\documentclass[twocolumn]{aastex631}
\usepackage{lineno}

\accepted{for publication in ApJ, October 7, 2024}

\begin{document}
 

\title{Multi-Observatory Research of Young Stellar Energetic Flares (MORYSEF): X-ray Flare Related Phenomena and Multi-epoch Behavior}

\correspondingauthor{Konstantin Getman}
\email{kug1@psu.edu}

\author[0000-0002-6137-8280]{Konstantin V. Getman}
\affiliation{Department of Astronomy \& Astrophysics \\
Pennsylvania State University \\ 
525 Davey Laboratory \\
University Park, PA 16802, USA}

\author[0000-0002-5077-6734]{Eric D. Feigelson}
\affiliation{Department of Astronomy \& Astrophysics \\
Pennsylvania State University \\ 
525 Davey Laboratory \\
University Park, PA 16802, USA}

\author[0000-0002-1566-389X]{Abygail R. Waggoner}
\affiliation{University of Virginia, Charlottesville, VA 22904, USA}

\author[0000-0003-2076-8001]{L. Ilsedore Cleeves}
\affiliation{University of Virginia, Charlottesville, VA 22904, USA}

\author[0000-0001-8694-4966]{Jan Forbrich}
\affiliation{Centre for Astrophysics Research, University of Hertfordshire, College Lane, Hatfield, AL10 9AB, UK}

\author[0000-0001-8720-5612]{Joe P. Ninan}
\affiliation{Department of Astronomy and Astrophysics, Tata Institute of Fundamental Research, Homi Bhabha Road, Colaba, Mumbai 400005, India}

\author[0000-0003-3061-4591]{Oleg Kochukhov}
\affiliation{Department of Physics and Astronomy, Uppsala University, Box 516, 75120 Uppsala, Sweden}

\author[0000-0003-4452-0588]{Vladimir S. Airapetian}
\affiliation{American University, 4400 Massachusetts Avenue NW, Washington, DC 20016, USA USA}
\affiliation{NASA/GSFC/SEEC, Greenbelt, MD 20771, USA}

\author[0000-0001-6010-6200]{Sergio A. Dzib}
\affiliation{Max-Planck-Institut fur Radioastronomie (MPIfR), Auf dem Hugel 69, 53121 Bonn, Germany}

\author[0000-0003-1413-1776]{Charles J. Law}
\altaffiliation{NASA Hubble Fellowship Program Sagan Fellow}
\affiliation{University of Virginia, Charlottesville, VA 22904, USA}

\author[0000-0003-1817-6576]{Christian Rab}
\affiliation{University Observatory, Faculty of Physics, Ludwig-Maximilians-Universitat Munchen, Scheinerstr. 1, D-81679 Munich, Germany}
\affiliation{Max-Planck-Institut für extraterrestrische Physik, Giessenbachstrasse 1, D-85748 Garching, Germany}

\begin{abstract}
The most powerful stellar flares driven by magnetic energy occur during the early pre-main sequence (PMS) phase. The Orion Nebula represents the nearest region populated by young stars, showing the greatest number of flares accessible to a single pointing of Chandra. This study is part of a multi-observatory project to explore stellar surface magnetic fields (with HET-HPF), particle ejections (VLBA), and disk ionization (ALMA) immediately following the detection of PMS super-flares with Chandra. In December 2023, we successfully conducted such a multi-telescope campaign. Additionally, by analyzing Chandra data from 2003, 2012, and 2016, we examine the multi-epoch behavior of PMS X-ray emission related to PMS magnetic cyclic activity and ubiquitous versus sample-confined mega-flaring. Our findings follow. 1) We report detailed stellar quiescent and flare X-ray properties for numerous HET/ALMA/VLBA targets, facilitating ongoing multi-wavelength analyses. 2) For numerous moderately energetic flares, we report correlations (or lack thereof) between flare energies and stellar mass/size (presence/absence of disks) for the first time. The former is attributed to the correlation between convection-driven dynamo and stellar volume, while the latter suggests the operation of solar-type flare mechanisms in PMS stars. 3) We find that most PMS stars exhibit minor long-term baseline variations, indicating the absence of intrinsic magnetic dynamo cycles or observational mitigation of cycles by saturated PMS X-rays. 4) We conclude that X-ray mega-flares are ubiquitous phenomena in PMS stars, which suggests that all protoplanetary disks and nascent planets are subject to violent high-energy emission and particle irradiation events.
\end{abstract}


\section{Introduction} \label{sec:intro}

\subsection{Pre-Main Sequence Stellar X-ray Emission} \label{sub:intro_xrays}

During the early phase of the stellar evolution, fully convective and fast rotating pre-main sequence (PMS) stars continue their descent along the Hayashi tracks on the Hertzsprung-Russell diagram, driven by gravitational contraction \citep[e.g.,][]{Getman22}. Many PMS stars in the first 1-5 Myr of their life are surrounded by protoplanetary disks \citep[e.g.,][]{Richert18}.

The time-averaged X-ray luminosity of a solar-mass PMS star is $10^3 - 10^4$ times that of the
present-day Sun \citep{Preibisch05}. High X-ray fluxes from a young Sun can be attributed to a convection-driven magnetic dynamo within the large convective PMS stellar interior, which is more efficient at generating strong magnetic fields than the tachoclinal dynamo operating in main sequence stars. The PMS dynamo generates increased magnetic fluxes, producing larger starspots, associated active regions, and X-ray emitting coronas \citep[e.g.,][]{Browning2008, Christensen2009,Getman2023}. Observed typical magnetic field strengths in active regions ($B_{spot}$) of PMS stars are often comparable to those of the current Sun, ranging from 1 to 5~kG \citep{Sokal2020}. However, the surface filling factors of these regions in PMS stars can exceed 80\%, in contrast to less than 10\% on the Sun \citep{Kochukhov2020}.

The bulk of the observed PMS coronal X-ray emission has two components: 1) quasi-continuous baseline emission (also known as ``quiescent'' or ``characteristic'' emission), likely arising from numerous small, unresolved X-ray flares \citep{Wolk05}, and 2) episodic large flares. Astrophysical modeling of the X-ray time-energy evolution of large flares from PMS stars, with and without disks, shows that many are driven by magnetic reconnection associated with gigantic magnetic coronal loops reaching altitudes of $(1-10)$~R$_{\star}$ \citep{Favata2005,Getman08b}. The time-averaged X-ray fluxes from young ($t<5$~Myr) PMS stars saturate with stellar rotation and age, possibly due to the high fractional coverage of surface magnetic active regions and/or associated extensive X-ray coronal structures \citep{Preibisch05,Getman22}.

Studies utilizing X-ray data from the {\it Chandra} Orion Ultradeep Project \citep[COUP;][]{Getman05} have examined the frequency and energetics of large PMS flares in the nearby ($d \sim 400$~pc) Orion Nebula \citep{Wolk05,Caramazza07,Colombo07}. {\it XMM} data \citep{Stelzer07} have been employed for the Taurus Molecular Cloud region ($d \sim 140$~pc). More recently, {\it Chandra} data have been used to investigate 40 more distant ($1 < d < 3$~kpc) and rich star-forming regions \citep{Getman2021}. These studies consistently show that the energy distribution of large PMS flares ($dN/dE_X \sim E_X^{-\alpha}$) follows a power-law with a slope of $\alpha \sim 2$ similar to what is observed in older stars and the contemporary Sun. However, in stark contrast to the Sun, solar-mass PMS stars exhibit flares that are millions of times more powerful and occur at a rate over a million times higher. The contemporary Sun is likely incapable of producing X-ray super-flares ($E_X > 10^{34}$~erg) and mega-flares ($E_X>10^{36}$~erg) that frequently occur in PMS stars \citep{Getman2021}.

Our current study is part of a multi-wavelength, multi-observatory project titled the Multi-Observatory Research of Young Stellar Energetic Flares (MORYSEF), aimed at investigating various aspects of PMS X-ray emission. These aspects include: 1) assessing the strength of PMS surface magnetic fields following a large X-ray flare, 2) investigating the effects of X-ray flares on disks' chemistry, 3) searching for flare-associated coronal mass ejections (CMEs), and 4) examining the multi-epoch behaviors of both characteristic and flare X-ray emission. 

To address aspects 1), 2), and 3), in December 2023, we initiated nearly simultaneous observations using the {\it Chandra} X-ray telescope, Habitable-zone Planet Finder (HPF) near-infrared instrument on the Hobby-Eberly Telescope (HET), Atacama Large Millimeter/submillimeter Array (ALMA), and Very Long Baseline Array (VLBA), focusing on young stars in the Orion Nebula star-forming region. {\it Chandra's} field of view encompasses large parts of the Orion Nebula Cluster (ONC) and the underlying Orion Molecular Cloud One (OMC-1). To address aspect 4), in addition to the X-ray observations in 2023, we re-analyzed earlier {\it Chandra} observations from 2003 (COUP observations), 2012, and 2016 of the same region.

\subsection{Scientific Objectives of our Multi-Telescope Campaign} \label{sub:intro_issues}

{\it 1) Assessing the strength of PMS surface magnetic fields following a large X-ray flare.} For normal stars, including main-sequence and PMS stars, a simple power law relationship has been established between X-ray luminosity and the integrated magnetic flux $\Phi = B \times A$, where $B$ is the surface magnetic field and $A$ is the surface area involved \citep{Pevtsov2003,Kirichenko2017,Getman2023}. This relationship, $\L_X \propto \Phi^m$ with m=1.0-1.5, holds over 12 orders of magnitude from the smallest solar magnetic structures to the continuous emission from PMS stars \citep[Figure~8 in][]{Getman2021b}. This relationship appears to indicate the universality of the propagation of magnetic flux from the stellar interior to the stellar surface, regardless of the nature of the underlying magnetic dynamo --- whether convection-driven in PMS stars or tachocline-driven in the current old Sun. 

However, a major discrepancy is seen with most powerful PMS super- and mega-flares with X-ray energies $E_X > 10^{35}$~erg \citep[Figure~8 in][]{Getman2021b}. Assuming equipartition of magnetic fields in modeled X-ray flaring coronal loops, the $L_X$ values are $10^3-10^4$ times stronger than the scaling relation predicts. Recent 3D-MHD flare calculations explain this with a steeper slope $m \simeq 3$ for powerful flares, but this requires magnetic field strength inside active regions of $B_{\text{spot}} \simeq 10-20$~kG \citep{Zhuleku2021}. These MHD calculations predict surface-averaged magnetic field  $<B>$ values much higher than those observed thus far in non-super-flaring PMS stars \citep{Sokal2020}. 

To search for unusually strong surface magnetic fields in PMS super-flaring stars --- greater than any observed on a solar or stellar surface --- we conducted a {\it Chandra} observation campaign in December 2023. Within $1-3$ days after the X-ray observations, we swiftly identified four super- and mega-flaring young stars in the Orion Nebula region, which exhibited properties suitable for near-infrared (NIR) Zeeman broadening measurements of magnetic-sensitive spectral lines, using data from the HET-HPF spectrograph. The Simbad names of these stars are COUP~881, COUP~1333, COUP~1424, and COUP~1463. The HPF spectra of these stars were obtained within a few weeks following the {\it Chandra} observations.

 In this paper, among other findings, we present the stellar and X-ray flare properties of these four stars. Detailed measurements of their surface magnetic fields utilising HPF data will be provided in a companion paper.

 {\it 2) Investigating the effects of X-ray flares on
disks' chemistry.} Powerful PMS emission has a substantial impact on young stellar environments, influencing protoplanetary disk and planet formation processes. X-rays, along with ultraviolet (UV) radiation, ionize, heat, and photoevaporate protoplanetary disks and primordial planetary atmospheres. A myriad of astrophysical processes may ensue in planet-forming disks: enhanced magneto-rotational instability affecting gas turbulence, viscosity, accretion and planetary migration; ion-molecular chemistry; dust grains sputtering; and disk corona ionization producing jets and winds.  Most of these X-ray ionization effects are theoretical \citep{Alexander2014,Glassgold2000,Glassgold2007,Owen19,Ercolano2021,Woitke2024}, but ionization products like [NeII] and variable HCO$^+$ have been observed in some disks \citep{Espaillat2023,Waggoner2023}. Moreover,  CMEs and stellar energetic particles (SEPs) associated with powerful X-ray flares might further intensify the chemical changes in disks and planets, as well as lead to the removal of disks and planetary atmospheres \citep{Airapetian2020,Hazra2022,Alvarado-Gomez2022,Airapetian23}.


Most theoretical studies assume steady X-ray irradiation of disks without considering the high-amplitude variations in flux and spectrum due to large X-ray flares \citep{Gorti2009,Owen2019}.  Recent time-dependent calculations show that disk ionization may respond to sudden X-ray flares \citep{Rab2017,Waggoner2022,Brunn2024}. HCO$^{+}$ is one of the most abundant molecular ions in protoplanetary disks and its optically thin isotopologue H$^{13}$CO$^{+}$ is generally bright. Rapid (on time scales of weeks) HCO$^{+}$ abundance changes are theoretically  predicted and can be explained by X-ray flares ionizing H$_{2}$ gas, producing H$_{3}^{+}$, which reacts with CO to produce HCO$^{+}$. Variable H$^{13}$CO$^{+}$ emission has indeed been detected by ALMA in the disk around the young star IM Lup, but no X-ray observations have been performed in the same epoch as the H$^{13}$CO$^{+}$ variability \citep{Cleeves2017}.

To investigate rapid changes in HCO$^{+}$ abundance due to the impact of large X-ray flares, we identified four young X-ray flaring stars in the Orion region (COUP~414, COUP~561, COUP~1174, and COUP~1333) during our December 2023 observation campaign, which have previously been measured for continuum dust and/or gas emission. We followed up on these stars with multiple ALMA observations over the following weeks.

In this paper, among other findings, we present the stellar and X-ray flare properties of these four stars. Detailed ALMA-based measurements of the temporal evolution of H$^{13}$CO$^{+}$ and/or HC$^{18}$O$^{+}$ in their disks will be provided in a companion paper.

{\it 3) Hunting for CMEs associated with large X-ray flares}. The impact of mega-flares $-$ X-rays and Extreme-UV (EUV), CME shocks and associated energetic particles $-$ on photoevaporation, erosion and chemistry of primordial planetary atmospheres is being studied via theoretical  models, but with a few empirical constraints \citep[e.g.,][]{Airapetian2020}.  It seems likely that steady (non-flare) stellar X-ray and ultraviolet (`XUV') radiation destroy primordial atmospheres of inner planets much faster ($<10^7$~yr) than hydrogen thermal hydrodynamic escape ($10^8$~yr), but the role of mega-flares is largely unknown. We might expect that PMS mega-flares produce unusually powerful CMEs which might ionize, heat, entrain and ablate young planetary atmospheres leading to even more rapid mass loss.

The signatures of CMEs have been indirectly inferred in over 20 active G, K, and M dwarf stars through the presence of Doppler shifts in emission and absorption lines \citep{Argiroffi2019,Namekata2022,Namekata2024}, as well as possible X-ray dimming and changes in X-ray absorption during flares \citep{Moschou2019,Veronig2021}. The latter effect has not been seen in super- and mega-flares produced by Orion Nebula PMS stars \citep{Favata2005,Getman08a}.

Direct detection of CMEs from PMS mega-flares may be possible in the radio band.  At early phases of CME development, nonthermal Type II, III and IV solar radio bursts are often seen at kHz-MHz frequencies \citep{Carley2020}, and at late phases when the CME occupies a large volume, thermal bremsstrahlung at higher frequencies of MHz-GHz is expected.  While numerous radio bursts in active M dwarfs have been detected, none could be clearly associated with CME-shock plasma emission \citep{Villadsen2019}. But the Orion Nebula PMS mega-flares are orders of magnitude more powerful than these dMe flares.  

CME emission associated with Orion Nebula mega-flares could be detected as a rise and fall of radio flux, possibly with frequency drift, over days following the X-ray mega-flares. This emission would be spatially resolved and displaced from the star on scales $\geq 1$~AU. During our December 2023 campaign, two VLBA observations were conducted in the 4 and 7 GHz bands within one to several days after the {\it Chandra} observations, capturing radio emission from several dozen X-ray super- and mega-flaring Orion Nebula stars.

The stellar and X-ray flare properties of these stars are reported in the current paper. Detailed VLBA-based astrometric analysis of CME-related emission due to potential wobbles in the radio positions of the flaring stars will be presented in a companion paper.

{\it 4) Examining the multi-epoch behaviors of both characteristic and flare X-ray emission.} Long-term (on time-scales of several to many years) variations in PMS characteristic X-ray emission could be linked to magnetic cyclic activity.

Dynamo-generated magnetic fields undergo complex evolution within stellar interiors. In the current Sun, such processes follow an 11-year cycle, causing observed variations in magnetic flux and associated sunspot numbers, the spatial distribution of sunspots, solar radiation, material ejection, and even the reversal of the Sun's magnetic polarity. Recent 3D HD and MHD simulations successfully reproduce many observed features of solar-type magnetic cyclical activity \citep{Kapyla2023}. Many old cool stars with tachocline dynamos are known to exhibit cyclic activity. For instance, a study examining Ca II H and K lines have revealed variations in chromospheric activity among more than 100 F1-M2 main-sequence stars, resulting in cyclic periods spanning years to decades \citep{Baliunas1995}. {\it XMM} and {\it Chandra} X-ray studies have reported observed variations in X-ray activity, supporting cycles lasting from 1.6 to 19 years in solar-type stars, including $\alpha~$Cen~A and B, 61~Cyg~A, HD~81809, $\iota$~Hor, and $\epsilon$~Eridani \citep{Ayres2023,Robrade2012,Orlando2017,Sanz-Forcada2019,Coffaro2020}.

For fully convective old M-dwarfs and young PMS stars, with their magnetic fields generated via convection-driven dynamos, simulations also predict magnetic cycles \citep{Yadav2016,Emeriau-Viard2017,Brown2020,Kapyla2021}. And indeed, magnetic cyclic activity has been observed in a dozen fully convective old M-dwarfs \citep{Wargelin2017,Ibanez2019,Ibanez2020,Irving2023}.

To date, there are no reported observations of long-term cyclic activity in young PMS stars. In the current paper, the characteristic X-ray emission fluxes of a few hundred Orion PMS stars are identified and compared across four different epochs. A sub-sample of several diskless stars exhibiting the highest variations of their X-ray characteristic levels is selected for future periodogram analysis. This analysis will involve nearly eighty additional archived {\it Chandra}-ACIS-HETG observations of the ONC, covering a time span of 20 years.

In parallel, examining the multi-epoch behavior of large PMS X-ray flares can shed light on the crucial question of whether such flaring is ubiquitous or confined to a subset of stars. In the former scenario, all protoplanetary disks and nascent planets around very young stars are subject to millions of violent high-energy irradiation events \citep{Getman2021}, while in the latter case, the significance of the impact of large flares on stellar environments will be much more limited. The current study addresses this question by providing evidence supporting the former scenario.

\subsection{Outline of the Paper} \label{sub:intro_outline}  Section \ref{sec:xray_data_reduction} details the process of {\it Chandra} data reduction and point source extraction.  Section~\ref{sec:stellar_sample} introduces the sample of young stars analyzed in this study. Section \ref{sec:flare_identification} presents the identification of large X-ray flares and baseline segments, along with the computation of associated energetics. Section \ref{sec:results} focuses on various scientific analyses, including investigation into the basic energy properties of large X-ray flares, description of stars targeted by HET/ALMA/VLBA and their {\it Chandra} X-ray flares, search for signs of magnetic cyclic activity, and examination of multi-epoch behaviors of large X-ray flares. Section \ref{sec:conclusions} summarizes our findings. Appendix \S~\ref{sec:chandra_me_changes} presents an evaluation of the temporal increase in the apparent {\it Chandra} X-ray median energies of young Orion Nebula stars due to {\it Chandra}'s sensitivity degradation.

\begin{deluxetable*}{cccc}
\tabletypesize{\normalsize}
\tablecaption{{\it Chandra} Observations \label{tab:chandra_observations}}
\tablewidth{0pt}
\tablehead{
\colhead{Obs. Id.} & \colhead{Exposure} & \colhead{PI} &
\colhead{Start Date} \\
\colhead{} & \colhead{(ksec)} & \colhead{} &  \colhead{(UT)} \\
\colhead{(1)} & \colhead{(2)} & \colhead{(3)} & \colhead{(4)} 
}
\startdata
4395 &	99.95 &	Feigelson & 2003-01-08 20:57:16 \\
3744 &	164.16 & Feigelson & 2003-01-10 16:16:36 \\
4373 &	171.47 & Feigelson & 2003-01-13 07:33:40 \\
4374 &	168.98 & Feigelson & 2003-01-15 23:59:34 \\
4396 &	164.55 & Feigelson & 2003-01-18 14:33:45 \\
3498 &	69.05  & Murray & 2003-01-21 06:09:25 \\
13637 &	27.24  & Forbrich & 2012-10-02 07:00:39 \\
14334 &	8.37 & Forbrich	& 2012-10-03 12:51:52 \\
14335 &	25.93 & Forbrich & 2012-10-04 08:19:03 \\
15546 &	17.83 & Forbrich & 2012-10-05 10:41:00 \\
17735 &	88.9 & Guenther & 2016-11-27 02:11:50 \\
26974 &	10.95 & Getman & 2023-12-17 04:26:31 \\
29116 &	13.91 & Getman & 2023-12-17 19:50:05 \\
29117 &	11.93 & Getman & 2023-12-18 04:53:32 \\
27059 &	9.96 & Getman & 2023-12-20 23:26:34 \\
29127 &	11.23 & Getman & 2023-12-21 06:10:36 \\
29128 &	9.47 & Getman & 2023-12-22 10:51:25 \\
\enddata 
\tablecomments{All observations were conducted in the VFAINT mode with the same aimpoint, located near the center of the ONC at ($\alpha$,$\delta$) = (05:35:16.7, -05:23:24.0) J2000.0. Columns 1-4: The {\it Chandra} observation ID, net exposure time, name of the observation's principal investigator, and start time of the observation in UT.}
\end{deluxetable*}

\begin{deluxetable*}{cccccccccccc}
\tabletypesize{\normalsize}
\tablecaption{Stellar Properties of 245 Stars Selected as the X-ray Brightest in the 2023 Epoch \label{tab:stellar_props}}
\tablewidth{0pt}
\tablehead{
\colhead{Src.} & \colhead{R.A.} & \colhead{Decl.} &
\colhead{SpT} & \colhead{$A_V$} & \colhead{$M$} & \colhead{$R$} & \colhead{$\alpha_{IRAC}$} & \colhead{$P_{rot}$} & \colhead{$v_{rot} sin(i)$} & \colhead{$ME$} & \colhead{$\log(L_{X})$} \\
\colhead{} & \colhead{(deg)} & \colhead{(deg)} &  \colhead{} & \colhead{(mag)} & \colhead{(M$_{\odot}$)} & \colhead{(R$_{\odot}$)} & \colhead{} & \colhead{(day)} & \colhead{(km~s$^{-1}$)} & \colhead{(keV)} & \colhead{(erg~s$^{-1}$)} \\
\colhead{(1)} & \colhead{(2)} & \colhead{(3)} & \colhead{(4)} & \colhead{(5)} & \colhead{(6)} & \colhead{(7)} & \colhead{(8)} & \colhead{(9)} & \colhead{(10)} & \colhead{(11)} & \colhead{(12)}
}
\startdata
COUP23 & 83.688341 & -5.417808 & K2 & 3.3 & 2.4 & 3.4 & -2.8 & 3.5 & 32.9 & 1.3 & 31.2 \\
COUP28 & 83.693401 & -5.408844 & M0 & 0.0 & 0.7 & 1.7 & -2.6 & \nodata & 22.7 & 1.4 & 30.8 \\
COUP43 & 83.703485 & -5.388333 & M1 & 2.1 & 0.9 & 2.2 & -2.5 & \nodata & 70.6 & 1.3 & 30.3 \\
COUP66 & 83.716877 & -5.411919 & M3.5e & 2.3 & 0.8 & 1.6 & -1.1 & \nodata & 13.8 & 1.4 & 30.2 \\
COUP67 & 83.717473 & -5.375534 & M2.5 & 2.3 & 0.9 & 2.7 & -1.8 & \nodata & 11.6 & 1.2 & 30.3 \\
COUP107 & 83.733255 & -5.386967 & K1e & \nodata & \nodata & \nodata & -2.8 & 17.4 & 10.3 & 1.3 & 31.3 \\
COUP108 & 83.733330 & -5.490630 & M1.5 & 2.4 & 1.1 & 1.8 & -2.8 & 1.9 & 27.9 & 1.4 & 30.4 \\
COUP112 & 83.735615 & -5.464103 & M2e & 1.4 & 0.8 & 1.2 & -1.4 & 6.5 & 15.4 & 1.3 & 30.4 \\
COUP115 & 83.736866 & -5.360111 & K7 & 8.5 & 2.0 & 2.5 & -2.6 & 3.4 & 19.0 & 1.7 & 30.6 \\
COUP124 & 83.740827 & -5.397962 & K8 & 3.5 & 2.3 & 3.4 & -1.6 & \nodata & 19.2 & 1.4 & 30.9 \\
\enddata 
\tablecomments{Only a few examples of the table entries are given here; the full machine-readable table for all 245 young stars is provided in the electronic edition of this paper. Columns 1-3: Source name, right ascension and declination for epoch J2000.0, from \citet{Getman05}. Column 4: Spectral type from \citet{Skiff2014}. Columns 5-7: Source's visual extinction, stellar mass, and radius are from Table 6 of \citet{Getman22}. These quantities were obtained utilizing near-IR photometry data and the PARSEC 1.2S PMS stellar evolutionary model of \citet{Bressan12,Chen14}. Column 8: Spectral energy distribution slope based on the {\it Spitzer}-IRAC data from \citet{Megeath2012}. Values of $\alpha_{IRAC}$ less than -1.9 and greater than or equal to -1.9 indicate diskless and disk-bearing stars, respectively. Columns 9-10: Stellar rotation information. Stellar rotation periods are obtained from the compilation of \citet{Davies2014}. Projected stellar rotation velocities, derived from APOGEE near-IR spectral data, are sourced from \citet{Kounkel2019}. Columns 11-12: X-ray median energy and time-averaged X-ray luminosity are based on the COUP data of \citet{Getman05}. For relatively faint X-ray stars, improved X-ray luminosities are from \citet{Getman22}.}
\end{deluxetable*}

\begin{figure*}
\centering
\includegraphics[width=0.95\textwidth]{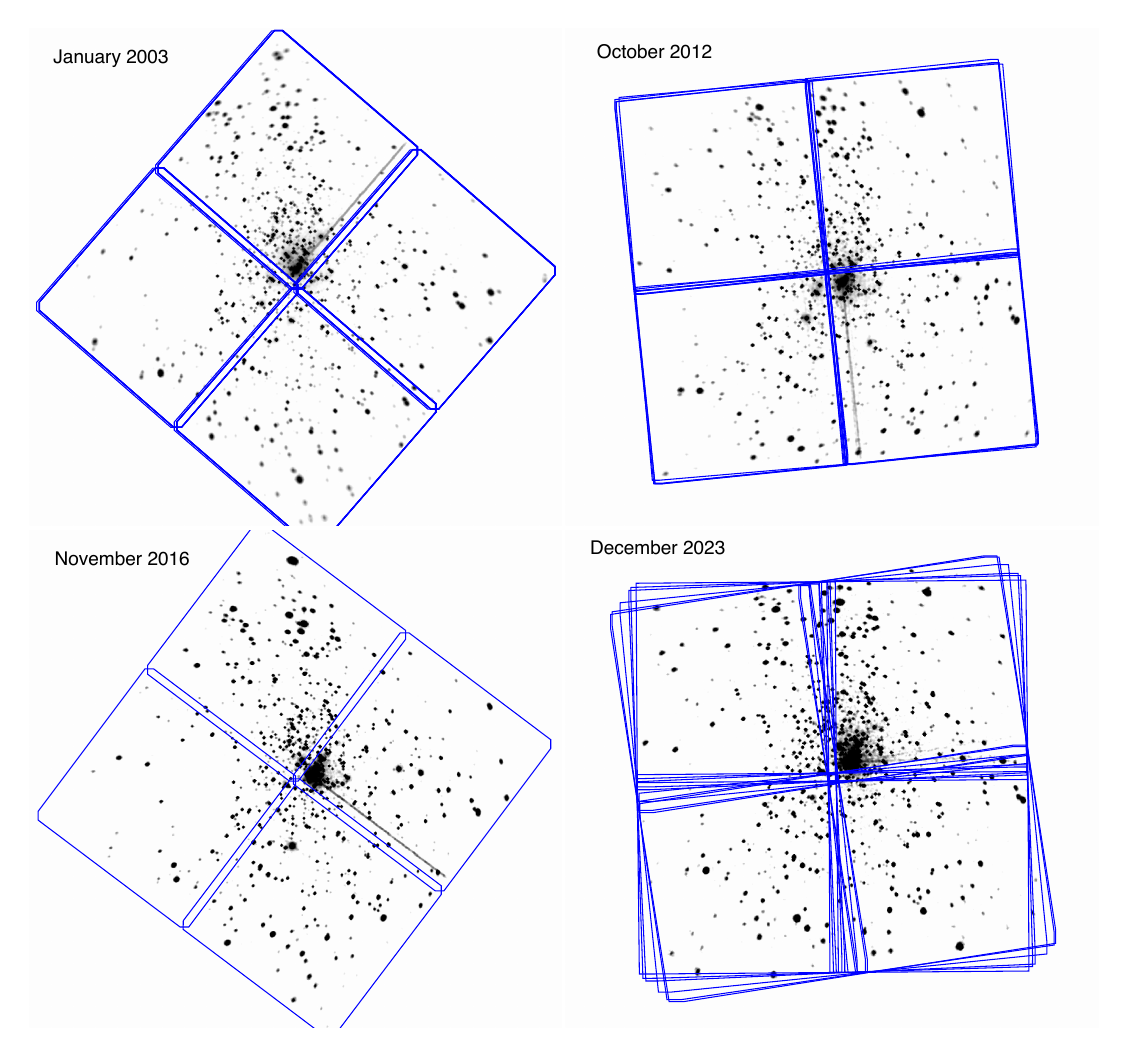}
\caption{Low-resolution, smoothed {\it Chandra}-ACIS-I images of the Orion Nebula region presented for each of the four distinct epochs of observations: January 2003, October 2012, November 2016, and December 2023. The images are given for the $(0.5-8)$~keV energy band. Each panel represents the combined view of multiple observations conducted during the corresponding epoch, with $17\arcmin \times 17\arcmin$ field of views for each observation outlined in blue. Due to variations in roll angles among observations within each epoch, the merged images sometimes display polygonal shapes rather than perfect squares.
} \label{fig:xray_images}
\end{figure*}

\begin{figure*}
\centering
\includegraphics[width=0.95\textwidth]{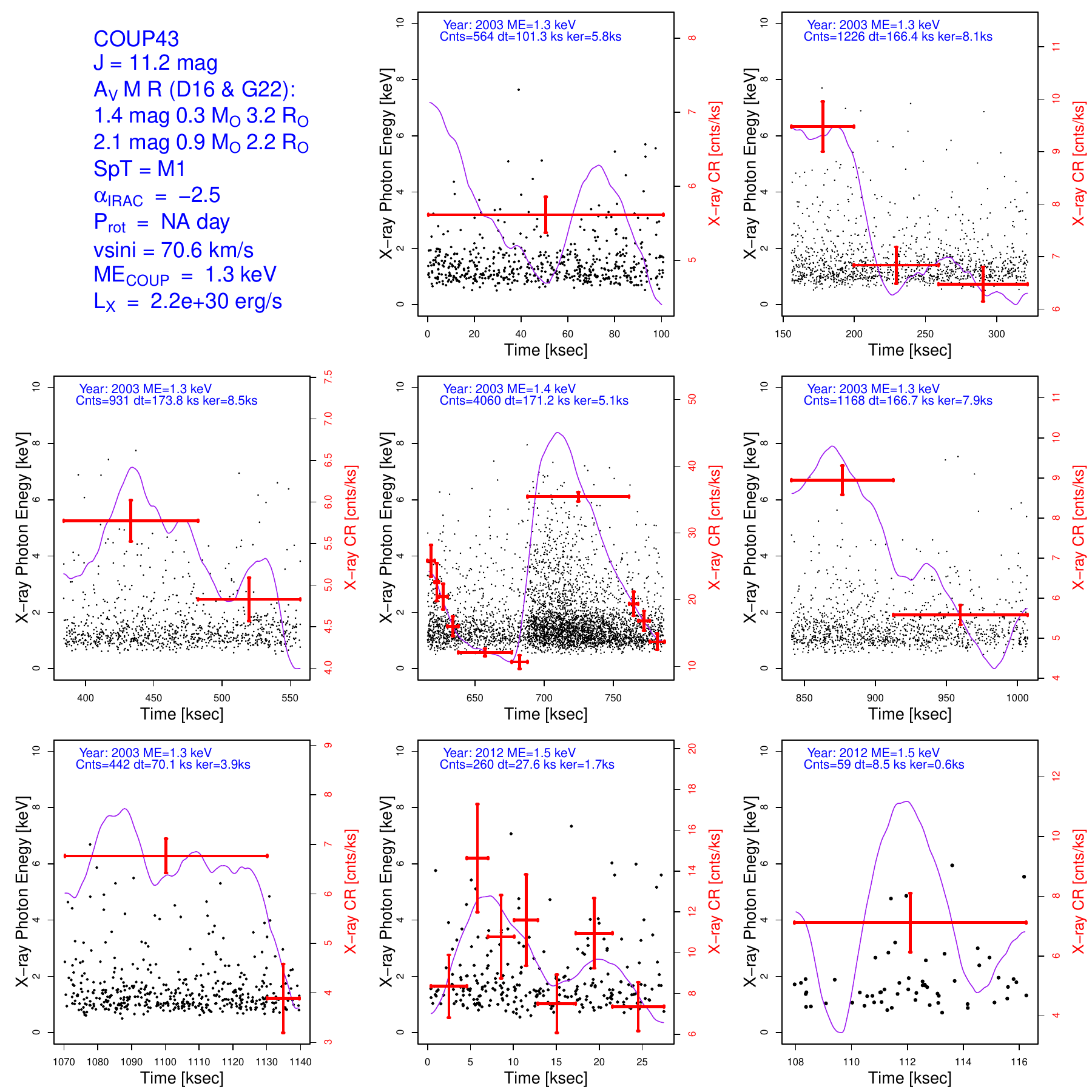}
\caption{Sample page from the flare atlas of Figure Set 2 featuring COUP~43. This atlas includes basic stellar properties extracted from Table~\ref{tab:stellar_props}, along with additional extinction-mass-radius estimates from \citet{DaRio2016}. It also showcases X-ray photon arrival diagrams and apparent lightcurves in count rates for each of the 17 {\it Chandra} observations listed in Table~\ref{tab:chandra_observations}. The abscissa time values are in kilo-seconds, and each of the 4-epoch observation sets begins at time 0~ksec. Black points on the photon arrival diagrams represent individual X-ray photons, with their energy values in keV shown on the left ordinates. Red segments and magenta curves denote parts of the lightcurves, with counts per kilo-second units displayed on the right ordinates. The red segments are outcomes from the Bayesian Block analysis, discussed further in the text. The magenta kernel density estimation (KDE) curves, also discussed in the text, provide visual insights into the lightcurve trends. The figure legends include observation epochs, total number of X-ray photons per observation, median energy of the photons, observation duration, and width of KDE kernel. One-$\sigma$ error bars for the Bayesian Blocks are based on the \citet{Gehrels86} statistic. The time zero points on the graphs for the epochs of 2003, 2012, 2016, and 2023 correspond to MJD values of 52647.89, 56202.31, 57719.11, and 60295.20 days, respectively. Refer to the next page for the continuation of the figure.} \label{fig:phot_arrival}
\end{figure*}

\begin{figure*}
\centering
\includegraphics[width=0.95\textwidth]{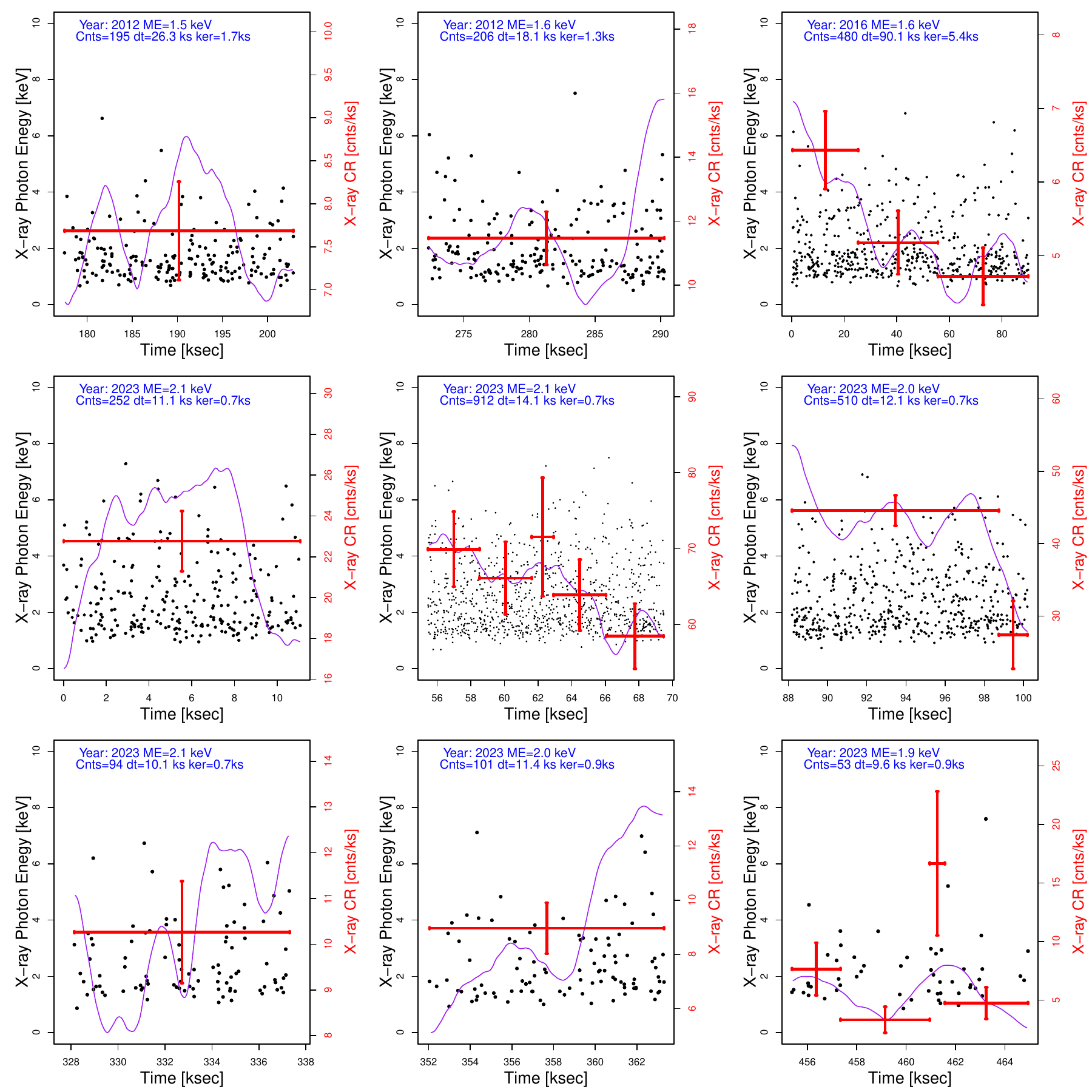}
\end{figure*}

\begin{figure*}
\centering
\includegraphics[width=0.95\textwidth]{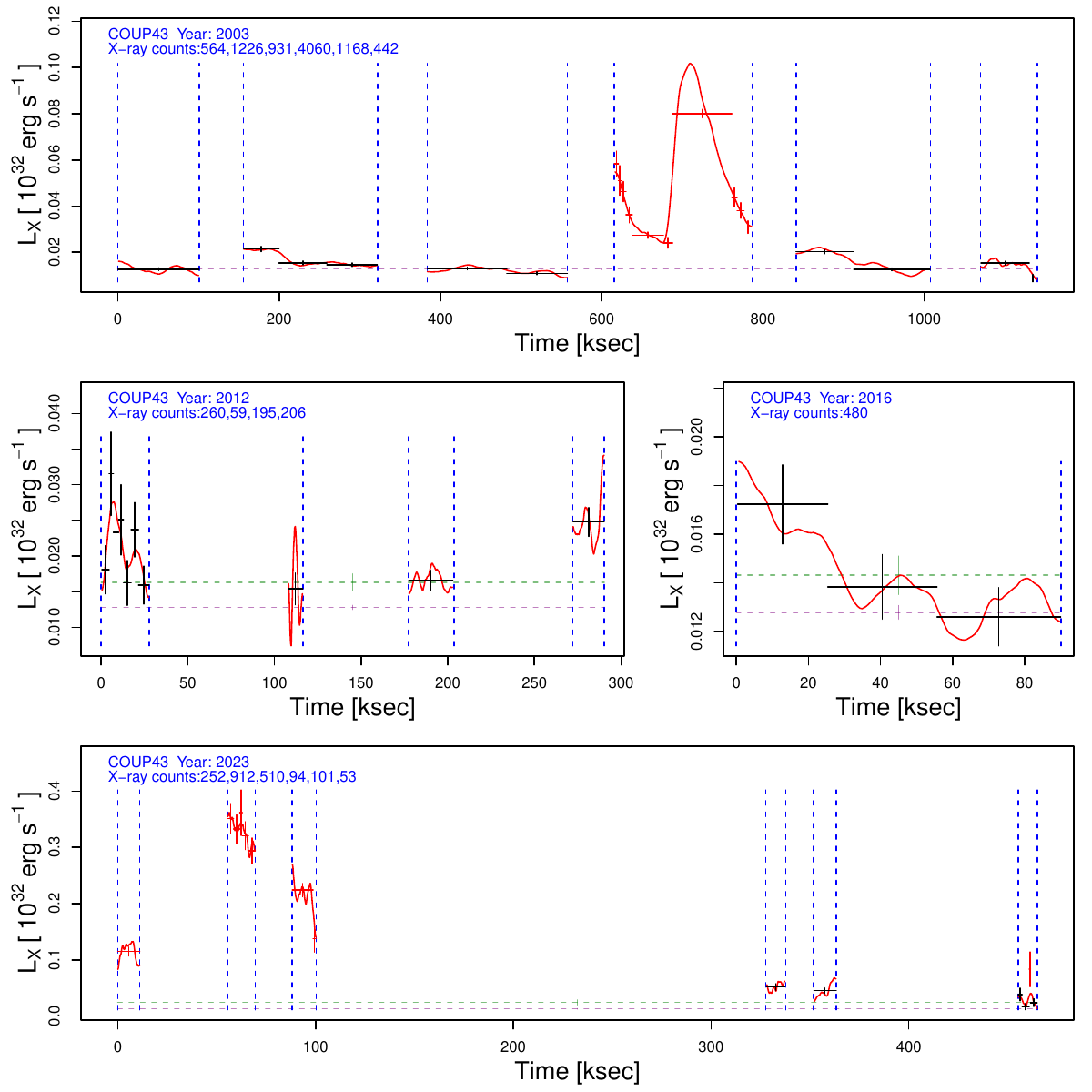}
\caption{Sample page from the flare atlas of Figure Set 3 featuring COUP~43. The atlas showcases X-ray lightcurves in units of intrinsic stellar X-ray luminosity, presented as four separate figure panels, each corresponding to one of the four-epoch observation sets. The blue dashed lines indicate the start and stop times for each of the 17 {\it Chandra} observations listed in Table\ref{tab:chandra_observations}. The KDE curves are shown in red. The red and black Bayesian Block segments represent cases of strong or mild-to-no-variability, respectively. The former denote short X-ray impulsive or long powerful flares, while the latter indicate characteristic emission or mild flare levels. The characteristic emission levels are also marked by dashed lines: magenta for the epoch-2003 and green for the 2012, 2016, and 2023 epochs. Standard 1-$\sigma$ confidence intervals (marked by error bars) for the Bayesian Blocks and characteristic levels are based on the \citet{Gehrels86} statistic. The figure legends include observation epochs and total number of X-ray photons per observation.} \label{fig:lc}
\end{figure*}

\section{X-ray Observations and Data Extraction} \label{sec:xray_data_reduction}

The X-ray data analyzed and discussed in this paper were acquired utilizing the {\it Chandra} X-ray Observatory \citep{Weisskopf2002}. The investigation involves 17 {\it Chandra} observations detailed in Table~\ref{tab:chandra_observations}, obtained with the imaging array of the Advanced CCD Imaging Spectrometer (ACIS-I). This array consists of four abutted $1024 \times 1024$~pixel$^2$ front-side–illuminated charge-coupled devices, covering approximately $17\arcmin \times 17\arcmin$ on the sky \citep{Garmire2003}. The observations were conducted during various epochs: six during January 2003 as part of the COUP project, four in October 2012, one in November 2016, and six in December 2023. Notably, the net {\it Chandra} exposure of the COUP project in 2003, totaling 838~ksec, is about 10 times longer than the net exposures of each of the three subsequent programs, which range between 67 and 89~ksec.

For the {\it Chandra} data reduction and analysis, CIAO v4.15.2 \citep{Fruscione2006}, CALDB v4.10.7, and HEASOFT
v6.31.1 \citep{Heasoft2014}  were utilized. 

During our December 2023 observation campaign, several data reduction tasks were employed to swiftly (within 1-2 days after a {\it Chandra} observation) analyze the new epoch-2023 {\it Chandra} data and identify stars with large X-ray flares, facilitating the designation of targets for immediate follow-up observations with HET, ALMA, and VLBA. For this purpose, basic {\it Chandra} data processing steps were conducted, including reprocessing from $L_1$ to $L_2$ event data, generating exposure maps, conducting a rapid search for bright point sources using the wavelet transform method \citep{Freeman2001}, correcting astrometry based on the COUP source catalog, and extracting data for several hundred sources using CIAO's {\it srcflux} tool. The source extraction process was simple, utilizing circular apertures for the sources while employing a single common background region. Subsequently, the resulting X-ray light curves were compared with the COUP source atlas \citep[Figure Set 12;][]{Getman05} to identify some young stars exhibiting prominent flares.

Later, we conducted a more thorough and consistent reduction and analysis of data from all 17 {\it Chandra} observations listed in Table~\ref{tab:chandra_observations}. Our approach to X-ray data reduction follows established procedures outlined in various previous {\it Chandra} studies of dense stellar clusters \citep[e.g.,][]{Kuhn2013a,Townsley2014,Getman17,Townsley2019,Getman22}.

Compared to the analysis conducted in December 2023, we employed more sophisticated yet time-intensive methods for source identification and extraction. Specifically, we utilized maximum likelihood image deconvolution with local point spread functions (PSFs) to enhance the resolution of closely spaced sources \citep{Broos2010}. Additionally, we employed the ACIS Extract (AE) software package, version 5658 2022-01-25 \citep{Broos2010,Broos2012}, to extract and characterize point sources from multi-ObsID ACIS data, based on local PSFs.

Through multiple iterations involving spatially crowded source candidates, AE generated optimal source and background extraction regions, refined source positions, distinguished between spurious and significant sources, and provided Poisson calculations for various X-ray properties of the sources, including net counts, median energies, and apparent photometric fluxes, among other parameters.

The original COUP catalog, based on the January 2003 {\it Chandra} ACIS-I data, contains 1616 X-ray point sources, with 1414 identified as young stellar members of the Orion Nebula region \citep{Getman05,Getman2005b}. Our current {\it automatic} re-analysis of these data identifies 1557 X-ray point sources, including 1348 COUP young stellar members. For the epochs in 2012, 2016, and 2023, AE produced lists of 1147, 1074, and 1029 X-ray point sources and recovers 975, 941, and 842 known COUP young stellar members, respectively. The multi-epoch {\it Chandra} images of the Orion Nebula region reveal hundreds of bright X-ray PMS stars (Figure~\ref{fig:xray_images}).

\section{PMS stellar sample} \label{sec:stellar_sample}
For the analysis of PMS X-ray emission presented in this paper, the initial multi-epoch sample of over 800 young stars from the Orion Nebula was narrowed down to a subset comprising relatively bright stars, each having at least 100 X-ray net counts in the epoch-2023 {\it Chandra} observations. This refined sub-sample consists of 245 young stars. Table~\ref{tab:stellar_props} provides an overview of their fundamental stellar characteristics, encompassing spectral types, source extinctions, stellar masses and radii, {\it Spitzer}-IRAC spectral energy distribution (SED) slopes, stellar rotation velocities, as well as COUP time-averaged X-ray median energies and X-ray luminosities. IRAC SED slopes serve as indicators of disk presence or absence. The X-ray median energy serves as a proxy for both coronal plasma temperature and absorbing gas column density \citep{Getman10}.

Among the 245 stars, estimates are available for 80\% of their spectral types, 95\% of their masses, 85\% of their IRAC SED slopes, and 65\% of their rotation quantities (either periods or projected rotation velocities). Regarding spectral types, 38\% are classified as M-type, 48\% as K-type, 9\% as G-type, and 5\% as A to O-type stars, respectively. The stellar masses in the sample range between 0.1 and 40~M$_{\odot}$, with a median value of 1.2~M$_{\odot}$. Similarly, stellar radii span from 1 to 9~R$_{\odot}$, with a median value of 2.5~R$_{\odot}$. And time-averaged COUP X-ray luminosities span from $8.7 \times 10^{28}$ to $1.6 \times 10^{33}$~erg~s$^{-1}$, with a median value of $3.1 \times 10^{30}$~erg~s$^{-1}$.  

Among stars with known IRAC SED slopes, 58\% are categorized as disk-bearing, while 42\% are identified as diskless. Disk-bearing stars typically exhibit systematically slower rotation due to star-disk magnetic coupling \citep{Matt2015,Garraffo2018}. Specifically, for diskless and disk-bearing stars, the median rotation velocities are 29 and 16~km~s$^{-1}$, respectively, and the median rotation periods are 5 and 9 days, respectively.

\section{X-ray lightcurves and identification of flare and characteristic segments} \label{sec:flare_identification}
Figures~\ref{fig:phot_arrival} and \ref{fig:lc} showcase sample pages from two flare atlases, displaying X-ray photon arrival diagrams and lightcurves. The complete atlases for all 245 stars are accessible as electronic Figure Sets \ref{fig:phot_arrival} and \ref{fig:lc}. The sample pages in Figures~\ref{fig:phot_arrival} and \ref{fig:lc} specifically highlight the lightly-absorbed, diskless star of type M1, COUP~43, a member of the ONC cluster. The star exhibited large X-ray flares during the epochs of 2003 and 2023.

The raw (unbinned) event data from each of the 17 {\it Chandra} observations, delivered by AE, underwent processing using the R-based {\it bayesian\_blocks.R} code\footnote{The code is available at \url{https://github.com/AlicePagano/ASPA-Project-Bayesian-Blocks-Algorithm-in-R}.}. This process yields an optimal segmentation of the data, identifying relevant time change-points. The code implements the Bayesian Blocks algorithm \citep{Scargle98,Scargle2013}, widely utilized in astronomy for flare identification \citep{Getman2021}. The algorithm partitions the data into blocks of varying widths, striking a balance between the goodness-of-fit of the blocks and the complexity of the partitioning. Goodness-of-fit is determined using the Cash maximum likelihood statistic, which is suitable for sparse Poisson-driven data points \citep{Cash1979}. The complexity of the partitioning is controlled by the prior distribution of the change-points in the form $\log(Prior) = 4 - \log(73.6 \times p0 \times N^{-0.5})$, where $p0$ is the scaling factor responsible for the algorithm's sensitivity, and $N$ is the number of data points in the time-series. In instances where the resulting number of blocks exceeded 10, we iteratively adjusted the slope of the prior distribution to provide a simpler solution.

The inferred Bayesian Blocks segments with their 1-$\sigma$ uncertainty ranges based on the \citet{Gehrels86} statistic are depicted in both flare atlases, as well as are listed in Table~\ref{tab:bb_segments}. These segments were utilized for the identification of X-ray characteristic and large flare emission levels, and calculation of the characteristic X-ray luminosity and energetics of large X-ray flares, as described below.

Additionally, the flare atlases include kernel density estimation (KDE) smooth curves shown in red\footnote{ These are constructed using R's function density(x, kernel=epanechnikov, cut=0).  The Epanechnikov kernel is optimal in a least squares sense \citep{Wand95}. It is well-established that the KDE method underperforms near data boundaries \citep{Hazelton2009}. To address this, we apply the data reflection technique, where data points are reflected at the boundaries, effectively extending the dataset beyond its original range. This approach helps mitigate boundary effects by providing additional density support near the edges.} It is important to note that these curves are intended solely as visual aids to illustrate potential trends in X-ray emission rise and decay. They do not, however, establish the statistical significance of related variability features.

For each source, the X-ray count rates corresponding to the Bayesian Blocks segments were translated into intrinsic X-ray luminosities in the $(0.5-8)$~keV energy band ($L_X$). This conversion was achieved using AE-based instrument-independent and point spread function (PSF)-corrected apparent X-ray photon fluxes ($F_{X,phot}$), expressed in units of photons~cm$^{-2}$~s$^{-1}$, and a singular conversion factor derived from $F_{X,phot}$ to $L_X$. The conversion factor was based on the time-averaged X-ray luminosity values extracted from the COUP dataset (Table~\ref{tab:stellar_props}). No adjustments were made for the effects of X-ray emission hardening during large X-ray flares. According to the calculations utilizing the Portable Interactive Multi-Mission Simulator (PIMMS\footnote{The tool is available on-line at \url{https://heasarc.gsfc.nasa.gov/cgi-bin/Tools/w3pimms/w3pimms.pl}.}), the X-ray hardening effect resulting from the heightened coronal plasma temperature (e.g., from 2 to 10 keV) during a significant X-ray flare \citep{Getman08a} will yield only a 5\% increase in the intrinsic X-ray luminosity for the same observed {\it Chandra} count rate.

Graphical output from Figure Sets \ref{fig:phot_arrival} and \ref{fig:lc} was examined and the Bayesian Blocks segments with the lowest $L_X$ values were chosen to represent the X-ray characteristic emission levels (see Column~9 in Table~\ref{tab:bb_segments}). The $L_X$ values of these segments, averaged across each observation epoch and weighted by the segments' durations, constitute the characteristic X-ray luminosities ($L_{X,char}$). These are detailed in Table~\ref{tab:flare_energetics} and depicted as horizontal dashed lines in Figure Set \ref{fig:lc}. 

Segments with X-ray luminosities more than three times higher than the characteristic luminosities were categorized as either short impulsive flares or large powerful flares. In Figure Set \ref{fig:lc}, these segments are highlighted in red. Subtracting the characteristic levels and integrating the X-ray luminosities of such segments over their durations yields flare energies, denoted as $E_{X,fl}$ (see Table~\ref{tab:flare_energetics}). For the observation epochs of 2012 and 2023, the $E_{X,fl}$ values associated with large X-ray flares represent only lower limits of the true energy values, given that the durations of individual 2012 and 2023 {\it Chandra} observations are shorter than the typical durations of large PMS X-ray flares, which range from 40 to 50 ksec \citep{Getman08a,Getman2021b}.

In the observations of 2003, 2012, 2016, and 2023, there are 222, 100, 44, and 157 identified X-ray flare events among the 245 young stars (Table~\ref{tab:flare_energetics}), respectively, with energies exceeding $\log(E_{X,fl}) > 34$~erg. The relatively low number of flares with inferred energy values {\it recorded} in 2016 can be attributed to the shorter total time span (including observation gaps), resulting in fewer reliable characteristic segments and thus making it difficult to accurately determine flare energies.

Lastly, segments with X-ray luminosities falling between the levels of flares and characteristic emission were categorized as ``mildly variable''. The combined energy from these segments, along with the flare segments, is denoted as $E_{X,var}$ and detailed in Table~\ref{tab:flare_energetics}.

In Figure Set \ref{fig:phot_arrival}, the stellar property legends include additional estimations for source extinction, stellar mass, and stellar radius from \citet{DaRio2016}, which are based on near-infrared data obtained with the APOGEE spectrograph and older PMS evolutionary models by \citet{Siess00}. When compared to the stellar properties from \citet{Getman2021}, as listed in Table~\ref{tab:stellar_props} and derived based on near-infrared photometry and newer PARSEC PMS evolutionary models, the masses and radii from Da Rio appear systematically lower and higher, respectively, by approximately 30\% and 15\%. This systematic difference primarily stems from variations in the utilized PMS models and the adoption of different color-$T_{eff}$ relations, as detailed in Appendix C of \citet{Getman2021}. We adopt the stellar properties from \citet{Getman2021} here simply because they are available for a larger number of stars.

Mainly due to the buildup of contamination on the optical blocking filters, the sensitivity of Chandra ACIS toward softer X-rays has degraded over time. As a result, the apparent Chandra X-ray median energies ($ME$) of sources have increased over time. The values of the apparent $ME$ provided in the panel legends of Figure~\ref{fig:phot_arrival} are especially useful for comparing these quantities among multiple Chandra observations taken during the same epoch. For instance, the $ME$ of COUP~43 reached a peak value of 1.4~keV during a large X-ray flare (observation \#4 for epoch 2003), while it remained constant at 1.3~keV during the other five Chandra observations, which primarily showed a characteristic level of X-ray emission in epoch 2003.

However, for inter-epoch comparisons of a source's $ME$, it is important to account for systematic shifts, which we derive in Appendix \S~\ref{sec:chandra_me_changes}.

\begin{deluxetable*}{ccccccccc}
\tabletypesize{\normalsize}
\tablecaption{Bayesian Blocks \label{tab:bb_segments}}
\tablewidth{0pt}
\tablehead{
\colhead{Src.} & \colhead{Year} & \colhead{Obs. \#} &
\colhead{$t_{start}$} & \colhead{$t_{stop}$} & \colhead{$Cnts$} & \colhead{$\log(L_{X,BB})$} & \colhead{$\sigma$\_$\log(L_{X,BB})$} & \colhead{Flag} \\
\colhead{} & \colhead{} & \colhead{} &  \colhead{(ksec)} & \colhead{(ksec)} & \colhead{(cnts)} & \colhead{(erg~s$^{-1}$)} & \colhead{(erg~s$^{-1}$)} & \colhead{} \\
\colhead{(1)} & \colhead{(2)} & \colhead{(3)} & \colhead{(4)} & \colhead{(5)} & \colhead{(6)} & \colhead{(7)} & \colhead{(8)} & \colhead{(9)}
}
\startdata
COUP28 & 2003 & 1 & 0.08 & 101.20 & 592 & 30.09 & 0.018 & c \\
COUP28 & 2003 & 2 & 156.07 & 322.27 & 865 & 30.04 & 0.015 & c \\
COUP28 & 2003 & 3 & 383.91 & 431.60 & 282 & 30.10 & 0.027 & n \\
COUP28 & 2003 & 3 & 431.60 & 435.83 & 99 & 30.70 & 0.046 & f \\
COUP28 & 2003 & 3 & 435.83 & 527.55 & 760 & 30.24 & 0.016 & n \\
COUP28 & 2003 & 3 & 527.55 & 531.65 & 90 & 30.67 & 0.048 & f \\
COUP28 & 2003 & 3 & 531.65 & 534.17 & 88 & 30.87 & 0.049 & f \\
COUP28 & 2003 & 3 & 534.17 & 557.53 & 2773 & 31.40 & 0.009 & f \\
COUP28 & 2003 & 4 & 615.71 & 625.68 & 3834 & 31.91 & 0.008 & f \\
COUP28 & 2003 & 4 & 625.68 & 636.53 & 3512 & 31.84 & 0.008 & f \\
COUP28 & 2003 & 4 & 636.53 & 651.55 & 2806 & 31.60 & 0.009 & f \\
COUP28 & 2003 & 4 & 651.55 & 677.62 & 2550 & 31.32 & 0.009 & f \\
COUP28 & 2003 & 4 & 677.62 & 714.48 & 2453 & 31.15 & 0.009 & f \\
COUP28 & 2003 & 4 & 714.48 & 716.02 & 62 & 30.93 & 0.059 & f \\
COUP28 & 2003 & 4 & 716.02 & 717.24 & 61 & 31.03 & 0.059 & f \\
COUP28 & 2003 & 4 & 717.24 & 786.90 & 2355 & 30.85 & 0.009 & f \\
COUP28 & 2003 & 5 & 841.07 & 841.57 & 20 & 30.93 & 0.108 & f \\
COUP28 & 2003 & 5 & 841.57 & 913.58 & 1777 & 30.72 & 0.011 & f \\
COUP28 & 2003 & 5 & 913.58 & 913.89 & 18 & 31.09 & 0.115 & f \\
COUP28 & 2003 & 5 & 913.89 & 1007.63 & 1850 & 30.62 & 0.010 & f \\
COUP28 & 2003 & 6 & 1069.89 & 1111.01 & 525 & 30.43 & 0.020 & n \\
COUP28 & 2003 & 6 & 1111.01 & 1139.98 & 293 & 30.33 & 0.026 & n \\
COUP28 & 2012 & 1 & 0.05 & 27.53 & 126 & 29.96 & 0.046 & c \\
COUP28 & 2012 & 2 & 107.85 & 116.29 & 36 & 29.93 & 0.082 & c \\
COUP28 & 2012 & 3 & 177.43 & 203.20 & 147 & 30.06 & 0.043 & n \\
COUP28 & 2012 & 4 & 272.38 & 290.26 & 81 & 29.96 & 0.056 & c \\
\enddata 
\tablecomments{Examples of the table entries are provided for the 2003 and 2012 {\it Chandra} observations of COUP~28. The full machine-readable table, containing 8691 entries for all four epochs of observations of all 245 young stars, is available in the electronic edition of this paper. Columns 1-3: Source name, epoch year, and the relative sequential numbering of {\it Chandra} observations within that epoch. Columns 4-9: Properties of the identified Bayesian Block segment, including start and stop times of the segment, X-ray counts, X-ray luminosity level and its uncertainty, and a flag indicating the segment's status: ``f'' for flare level, ``c'' for characteristic level, and ``n'' for mild variability between the flare and characteristic levels.}
\end{deluxetable*}


\begin{deluxetable*}{cccccccc}
\tabletypesize{\normalsize}
\tablecaption{Flare Energetics \label{tab:flare_energetics}}
\tablewidth{0pt}
\tablehead{
\colhead{Src.} & \colhead{Year} & \colhead{$\log(L_{X,char})$} &
\colhead{$\sigma$\_$\log(L_{X,char})$} & \colhead{$\log(E_{X,fl})$} & \colhead{$\sigma$\_$\log(E_{X,fl})$} & \colhead{$\log(E_{X,var})$} & \colhead{$\sigma$\_$\log(E_{X,var})$} \\
\colhead{} & \colhead{} & \colhead{(erg~s$^{-1}$)} &  \colhead{(erg~s$^{-1}$)} & \colhead{(erg)} & \colhead{(erg)} & \colhead{(erg)} & \colhead{(erg)} \\
\colhead{(1)} & \colhead{(2)} & \colhead{(3)} & \colhead{(4)} & \colhead{(5)} & \colhead{(6)} & \colhead{(7)} & \colhead{(8)} 
}
\startdata
COUP23 & 2003 & 31.09 & 0.012 & \nodata & \nodata & 36.47 & 0.015 \\
COUP23 & 2012 & 30.89 & 0.016 & \nodata & \nodata & 35.34 & 0.027 \\
COUP23 & 2016 & 30.90 & 0.010 & \nodata & \nodata & \nodata & \nodata \\
COUP23 & 2023 & 30.90 & 0.025 & \nodata & \nodata & 35.13 & 0.066 \\
COUP28 & 2003 & 30.06 & 0.012 & 36.67 & 0.003 & 36.69 & 0.003 \\
COUP28 & 2012 & 29.96 & 0.033 & \nodata & \nodata & 33.80 & 0.238 \\
COUP28 & 2016 & 29.97 & 0.052 & \nodata & \nodata & 34.57 & 0.114 \\
COUP28 & 2023 & 29.78 & 0.065 & 34.28 & 0.088 & 34.35 & 0.086 \\
COUP43 & 2003 & 30.11 & 0.009 & 35.84 & 0.010 & 35.91 & 0.010 \\
COUP43 & 2012 & 30.21 & 0.031 & \nodata & \nodata & 34.43 & 0.093 \\
COUP43 & 2016 & 30.16 & 0.024 & \nodata & \nodata & \nodata & \nodata \\
COUP43 & 2023 & 30.38 & 0.076 & 35.88 & 0.014 & 35.91 & 0.014 \\
\enddata 
\tablecomments{Examples of the table entries are provided for a few stars. The full machine-readable table, containing 980 entries for all four epochs of observations of all 245 young stars, is available in the electronic edition of this paper. Columns 1-2: Source name and epoch year. Columns 3-4: X-ray characteristic (baseline) level and its 1-$\sigma$ uncertainty. Columns 5-6: X-ray energy (and its 1-$\sigma$ uncertainty) of all large and/or impulsive flares (i.e., including only Bayesian Blocks marked in red in Figure~\ref{fig:lc}). Columns 7-8: The total X-ray energy (and its 1-$\sigma$ uncertainty) for the sum of large flares and mild variable events, i.e., all Bayesian Blocks segments that are above the characteristic level. These include all Bayesian Blocks marked in red in Figure~\ref{fig:lc}, as well as some Bayesian Blocks marked in black, which are identified as ``mild variable''.}
\end{deluxetable*}

\begin{figure*}
\centering
\includegraphics[width=0.8\textwidth]{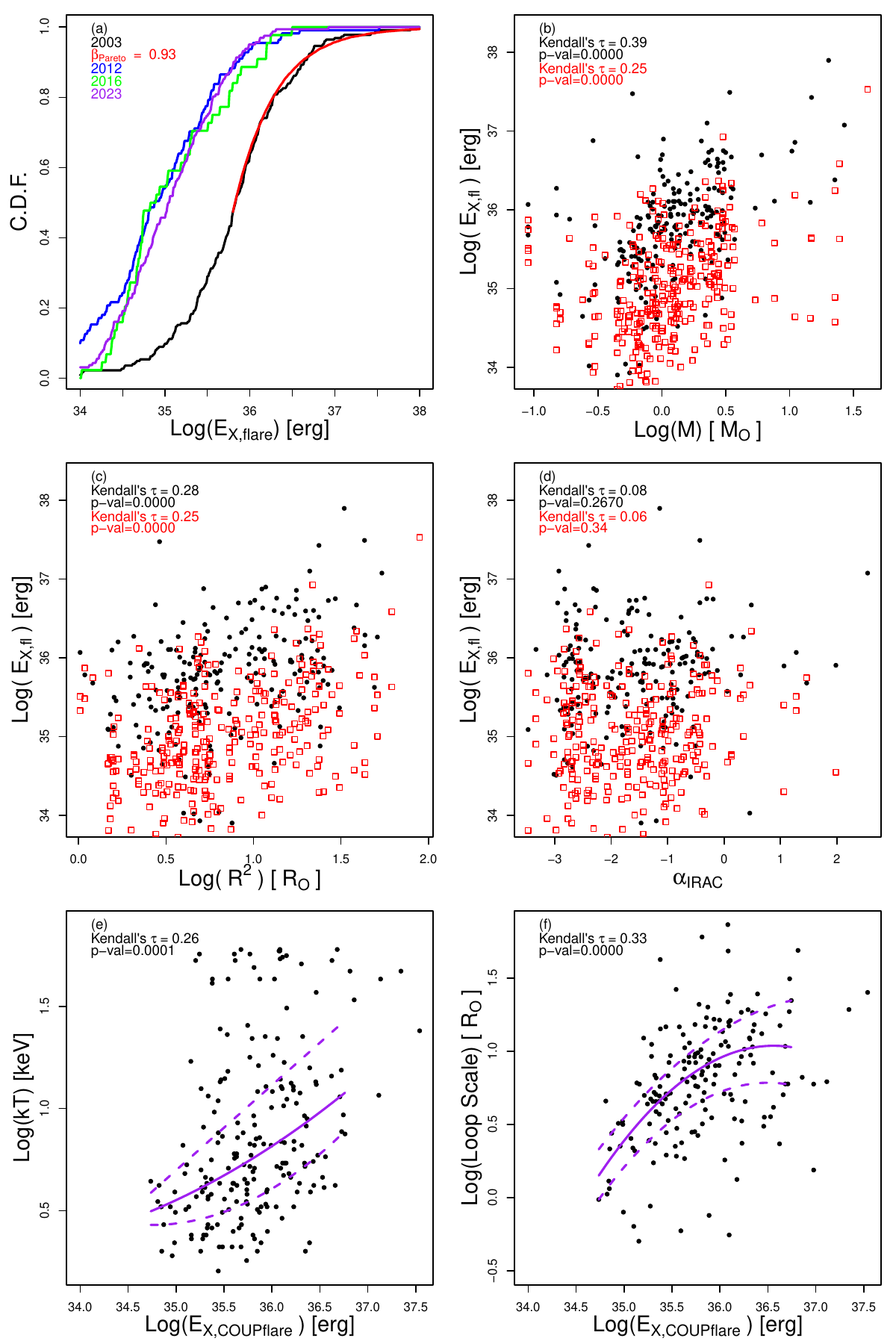}
\caption{(a) Cumulative distribution functions (CDFs) illustrating the X-ray flare energy distributions for the 2003 epoch (in black) and later (2012, 2016, and 2023) epochs (in color). The bright end of the 2003 epoch CDF is fitted with the Pareto (power law) function (red), with a slope of $\beta = 0.91$. (b-d) Scatter plots depicting the energies of X-ray flares observed during the 2003 epoch (in black) and subsequent epochs (in red), plotted against stellar mass, stellar surface area, and SED IRAC slope. Corresponding legends present the $\tau$ coefficients and $p$-values obtained from Kendall's $\tau$ test. (e-f) Depiction of guiding trends: the peak plasma temperature and flare loop scale as functions of X-ray flare energy, based on the COUP flare sample and flare quantities from \citet{Getman08a}. The purple curves represent spline fits to the 25\%, 50\%, and 75\% quartiles of the distributions, created using R's Constrained B-Splines {\it cobs} function \citep{Ng2007}.} \label{fig:flare_basics}
\end{figure*}
\clearpage

\section{Results} \label{sec:results}  

\subsection{Basic energy properties of detected large X-ray flares} \label{sec:flares_props}
Figures~\ref{fig:flare_basics}(a-d) depict the energy distributions of the large X-ray flares identified in this study (Table~\ref{tab:flare_energetics}). They also illustrate the positive correlations between these flare energies and stellar mass and size, respectively, and no correlations between flare energies and the presence or absence of protoplanetary disks.

These observed patterns in the distribution of flare energies align with findings from previous studies of large PMS X-ray flares, as discussed below.

Notably, Figure~\ref{fig:flare_basics}(a) illustrates that the energy distributions of flares observed during {\it Late} epochs (2012, 2016, and 2023) exhibit consistency among themselves but are systematically shifted towards lower energies compared to those observed during the {\it Early} epoch (2003). This discrepancy can be attributed to the decreasing frequency of stellar flares with increasing energy \citep{Getman2021}. Consequently, the approximately tenfold longer observations during the Early epoch capture a higher number of relatively rare, more powerful flares. Note that our simple flare identification algorithm (see \S~\ref{sec:flare_identification}) treats all flares detected within a single epoch as a single flare event. Thus, the presence of extremely long and powerful flares predominates the observed flare energy distribution for the Early epoch observations. Therefore, there appears to be an apparent lack of lower flare energies in the energy distribution curve for the Early epoch, although this absence is not intrinsic.

Figure~\ref{fig:flare_basics}(a) further illustrates that the upper range of flare energy distribution during the Early epoch can be adequately modeled by a Pareto function \citep{Arnold1983pareto}, with the Pareto slope $\beta$ approximately ranging between 0.9 and 1.

For a detailed examination of the properties of the Pareto function, refer to \citet{Getman2021}. Following \citet{Getman2021}, we conducted multiple trials using the Early epoch flare energy data with various cutoff X-ray flare energies ($E_{X,cutoff}$). We applied the maximum likelihood estimation of the Pareto slope \citep[formula (2) in][]{Getman2021}, generated the Pareto distribution CDF using the estimated slope, and performed the Anderson-Darling (AD) test to compare the unbinned and truncated observed data ($>E_{X,cutoff}$) with the Pareto CDF. The AD p-values were then evaluated to determine whether $E_{X,cutoff}$ corresponds to the flare energy completeness limit. We found that AD p-values were all $<0.01$ for $E_{X,cutoff}<10^{35.7}$~erg, indicating that the underlying Pareto distributions do not adequately represent the data. Conversely, for $E_{X,cutoff} \geq 10^{35.7}$~erg, AD p-values were above 0.1, suggesting that any of these cases can be used as reasonable solutions for the Pareto slopes. Statistical uncertainties on the Pareto slopes were estimated via bootstrap resampling. For example, with $\log(E_{X,cutoff})=[35.7,35.8,35.9,36.0]$~erg, the truncated data sets contain $N=[137,117,101,80]$ flare energy points, yielding AD p-values of $[0.14,0.22,0.3,0.22]$ and Pareto slopes of $\beta=[0.88 \pm 0.09, 0.93 \pm 0.09, 0.98 \pm 0.12, 1.00 \pm 0.13]$, respectively. Figure 4a illustrates the case of $\log(E_{X,cutoff})=35.8$~erg.

In the context of the energy distribution form ($dN/dE_X \sim E_X^{-\alpha}$), where $\alpha = \beta + 1$, this implies an $\alpha$ slope in the range of roughly  1.9 to 2.0. Remarkably, this slope value aligns with the flare energy distribution observed in various types of normal stars, including PMS stars, older stars, and even the Sun \citep{Caramazza07,Colombo07,Stelzer07,Getman2021}. The consistency in slope indicates a common flare production mechanism across all these stars.

Figures~\ref{fig:flare_basics}(b,c) demonstrate statistically strong positive correlations (with p-values from Kendall's $\tau$ test of $<0.0001$) between flare energy and stellar mass/size for both Early and Late epochs. These and similar correlations, previously illustrated in Figure~6 of \citet{Getman2021b}, can be attributed to the nature of magnetic fields in fully convective PMS stars. Unlike solar-type main sequence stars, where magnetic fields are generated in a thin tachocline layer (known as the $\alpha\Omega$ dynamo), fully convective PMS stars possess a distributed $\alpha^2$ dynamo throughout their interior \citep{Browning2008,Christensen2009,Kapyla2023}. In these fast-rotating stars, the strength of the surface magnetic field is determined by the kinetic energy of convective flows within the entire stellar interior, rather than rotation \citep{Christensen2009,Reiners2009,Reiners2010,Reiners2022}. Consequently, larger PMS stellar volumes (and hence mass and surface area) result in stronger magnetic fluxes, larger coronal active regions, and three-dimensional extensions of X-ray emitting structures \citep{Getman22,Getman2023}. This not only explains the correlations between flare energy ($E_{X,fl}$) and mass for large flares but also the positive correlations between time-averaged stellar X-ray luminosities and mass \citep{Feigelson93,Preibisch05,Telleschi07,Getman22}. Remarkably, our current results on flare energetics show, complementing previous findings of \citet{Getman2021b}, that even smaller X-ray flares ($\log(E_{X,fl}) < 35$~erg) positively correlate with stellar mass/size, marking as the first for this type of observation.

Figure~\ref{fig:flare_basics}(d) demonstrates that the energetics of X-ray flares show no correlation with the presence or absence of disks. This observation aligns with previous findings by \citet{Getman08b,Getman2021b}, who argued that the properties of large X-ray flares in PMS stars remain consistent regardless of the presence or absence of protoplanetary disks, as inferred from infrared photometry. This supports the solar-type model of flaring coronal magnetic loops, wherein both footpoints are anchored in the stellar surface rather than having one footpoint at the inner rim of the disk. Notably, our current flare energetics results extend this finding to less powerful X-ray flares ($\log(E_{X,fl}) < 35$~erg), marking the first time such disk independence has been demonstrated.

No correlation was found between PMS X-ray flare energies and stellar rotation rates (graph not shown). This lack of correlation aligns with expectations that the kinetic energy of convective flows and the generation of strong magnetic fields in fast-rotating PMS stars exhibit little to no dependence on rotation \citep{Christensen2009,Reiners2022}. Consistent with these findings, the time-averaged X-ray luminosities of fully convective PMS stars are also observed to be independent of rotation rates \citep{Preibisch05,Alexander2012}.

\begin{figure*}
\centering
\includegraphics[width=0.95\textwidth]{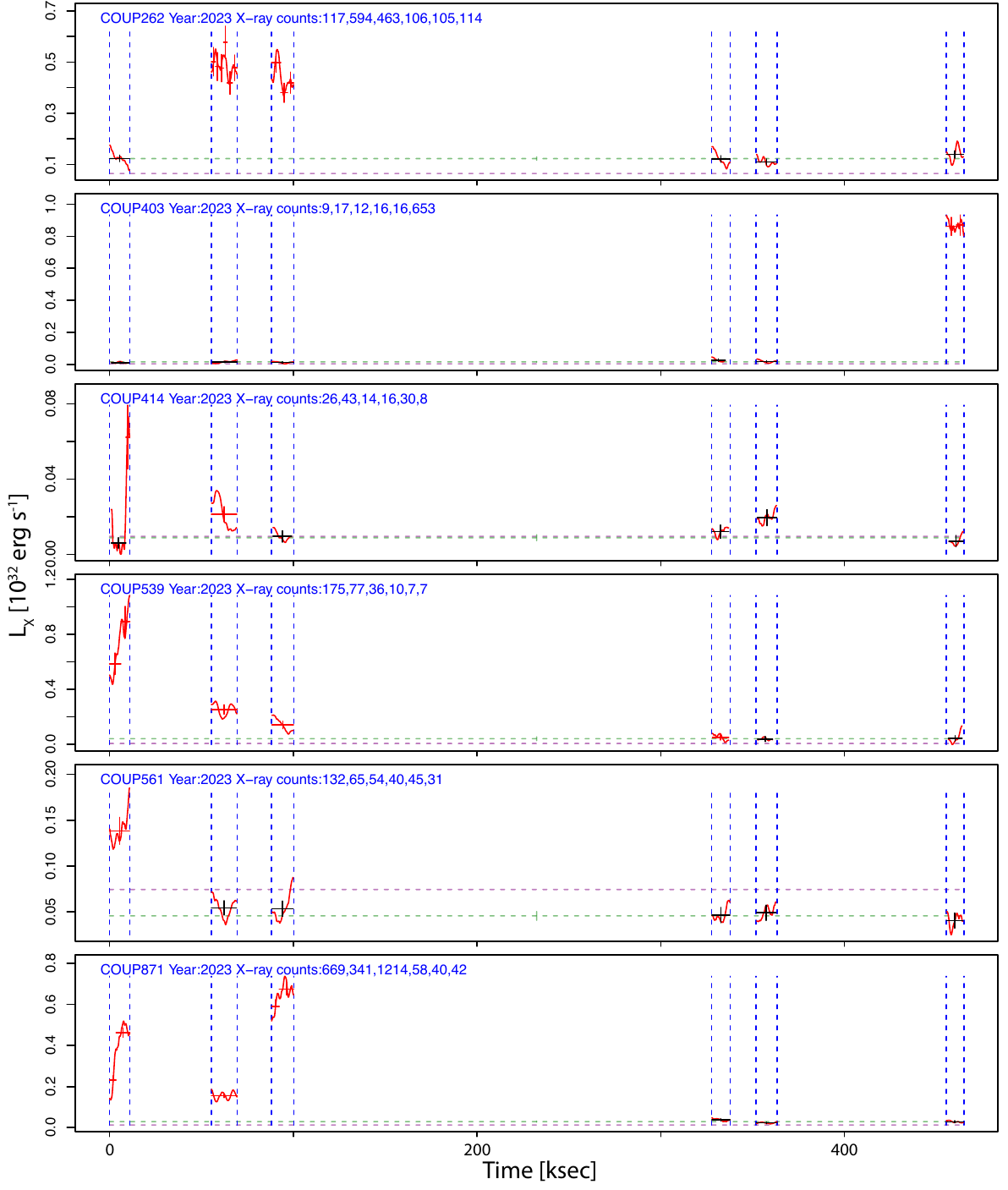}
\caption{Examples of X-ray light curves presented in units of intrinsic stellar X-ray luminosity for select targets observed with HET-HPF, ALMA, and VLBA. For further details, consult the caption of Figure~\ref{fig:lc}. Refer to the next page for the continuation of the figure.} \label{fig:lc_12_examples}
\end{figure*}

\begin{figure*}
\centering
\includegraphics[width=0.95\textwidth]{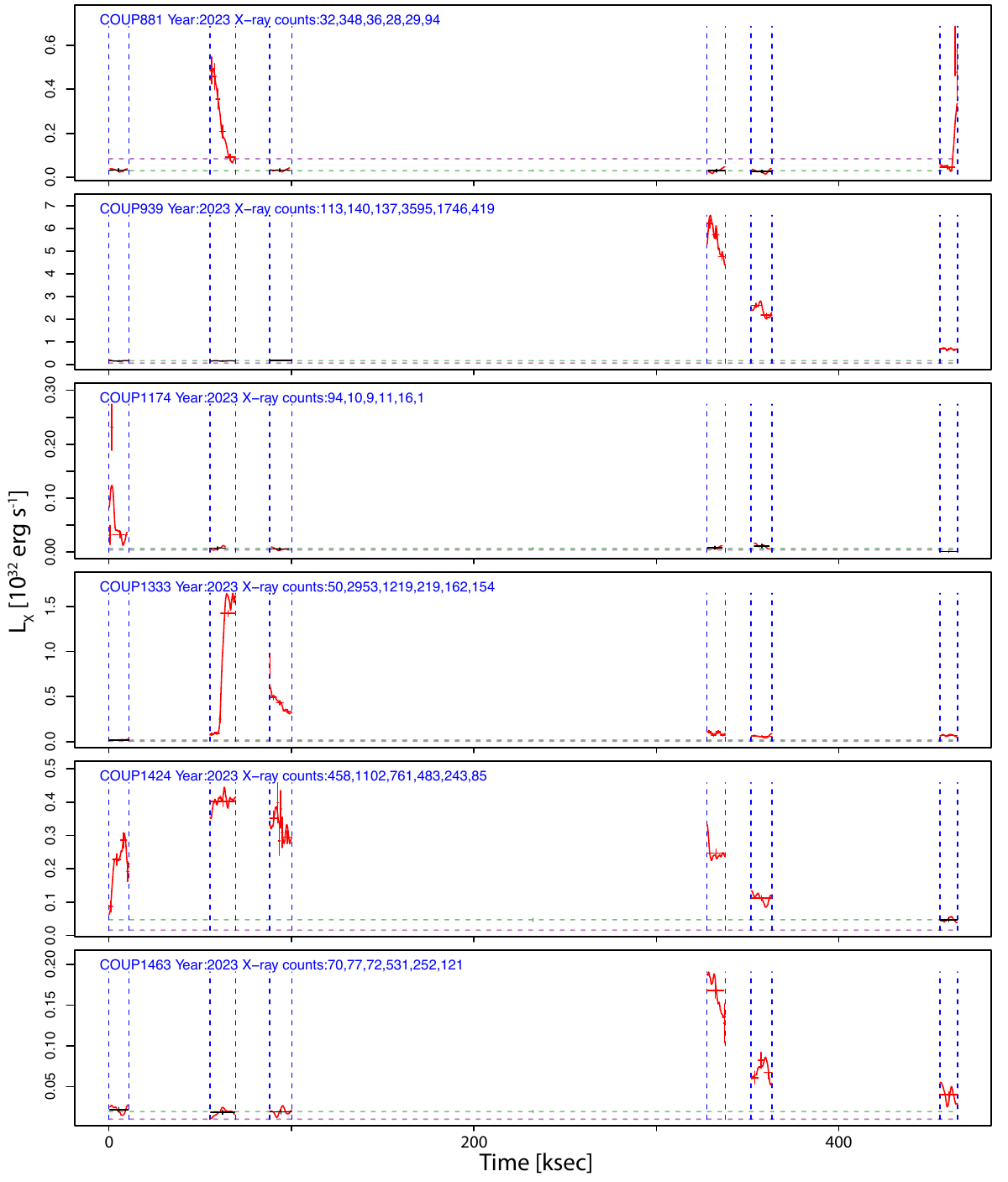}
\end{figure*}

\begin{deluxetable*}{cccccccccccc}
\tabletypesize{\normalsize}
\tablecaption{VLBA, ALMA, and HET-HPF Targets Observed in December 2023 \label{tab:vlba_alma_het_targets}}
\tablewidth{0pt}
\tablehead{
\colhead{Src.} & \colhead{$\log(E_{X,fl})$} & \colhead{$\sigma$\_$\log(E_{X,fl})$} &
\colhead{SpT} & \colhead{$A_V$} & \colhead{$M$} & \colhead{$R$} & \colhead{$\alpha_{IRAC}$} & \colhead{$P_{rot}$} & \colhead{$v_{rot} sin(i)$} & \colhead{$ME$} & \colhead{$\log(L_{X})$} \\
\colhead{} & \colhead{(erg)} & \colhead{(erg)} &  \colhead{} & \colhead{(mag)} & \colhead{(M$_{\odot}$)} & \colhead{(R$_{\odot}$)} & \colhead{} & \colhead{(day)} & \colhead{(km~s$^{-1}$)} & \colhead{(keV)} & \colhead{(erg~s$^{-1}$)} \\
\colhead{(1)} & \colhead{(2)} & \colhead{(3)} & \colhead{(4)} & \colhead{(5)} & \colhead{(6)} & \colhead{(7)} & \colhead{(8)} & \colhead{(9)} & \colhead{(10)} & \colhead{(11)} & \colhead{(12)}
}
\startdata
COUP939 & 36.93 & 0.007 & K8e & 10.8 & 3.0 & 4.7 & -0.3 & \nodata & \nodata & 1.8 & 30.9 \\
COUP1035 & 36.30 & 0.022 & \nodata & 18.5 & 3.0 & 4.9 & -1.9 & \nodata & \nodata & 2.3 & 31.1 \\
COUP1333 & 36.27 & 0.008 & em,M0.5 & 1.1 & 0.8 & 2.2 & -0.9 & 9.0 & 13.8 & 1.2 & 30.1 \\
COUP621 & 36.25 & 0.024 & \nodata & 35.6 & 22.6 & 6.1 & \nodata & \nodata & \nodata & 3.6 & 30.9 \\
COUP1298 & 36.16 & 0.023 & \nodata & 28.2 & 2.2 & 4.9 & -0.5 & \nodata & \nodata & 3.4 & 29.7 \\
COUP871 & 36.12 & 0.011 & M0.5 & 3.4 & 0.8 & 1.5 & \nodata & \nodata & 8.3 & 1.5 & 30.3 \\
COUP1424 & 36.11 & 0.011 & M1 & 2.1 & 0.9 & 1.9 & -2.8 & 10.6 & 12.2 & 1.3 & 30.3 \\
COUP539 & 36.07 & 0.036 & \nodata & \nodata & \nodata & \nodata & \nodata & \nodata & \nodata & 4.0 & 30.2 \\
COUP262 & 35.95 & 0.020 & K5 & 11.4 & 3.0 & 4.7 & -2.6 & 3.8 & 27.5 & 1.9 & 31.1 \\
COUP444 & 35.92 & 0.028 & \nodata & 28.2 & 0.5 & 2.5 & -1.0 & \nodata & \nodata & 3.0 & 30.1 \\
COUP403 & 35.91 & 0.022 & K?,M2.7 & 2.8 & 0.3 & 2.3 & -0.3 & \nodata & \nodata & 3.2 & 30.1 \\
COUP43 & 35.88 & 0.014 & M1 & 2.1 & 0.9 & 2.2 & -2.5 & \nodata & 70.6 & 1.3 & 30.3 \\
\enddata 
\tablecomments{A subset of 81 stars exhibiting prominent X-ray flares in December 2023 and observed with either VLBA, ALMA, or HET-HPF. The COUP~881, COUP~1333, COUP~1424, and COUP~1463 stars have been observed by HET-HPF. The COUP~414, COUP~561, COUP~1174, and COUP~1333 have been observed by ALMA. And all but one (COUP~561) of the stars listed in this table have been observed by VLBA. These targets are listed in descending order of their X-ray flare energy values. Examples of table entries are provided here; the full machine-readable table for all 81 young stars is available in the electronic edition of this paper. Column 1: Source name. Columns 2-3: Energy values of large X-ray flares detected in December 2023 (from Table~\ref{tab:flare_energetics}). Columns 4-12: Stellar properties including spectral type, source's visual extinction, stellar mass and radius, SED IRAC slope, stellar rotation period and projected velocity, X-ray median energy, and time-averaged X-ray luminosity (all from Table~\ref{tab:stellar_props}).
}
\end{deluxetable*}

\subsection{Stars and their X-ray flares for HET, ALMA, VLBA studies} \label{sec:stars_flares_for_het_alma_vlba}
In this section, we outline the X-ray flaring stars, and the properties of their flares, identified in the December 2023 {\it Chandra} observations and subsequently observed with HET-HPF, ALMA, and VLBA.

Due to the nature of the December 2023 {\it Chandra} data, consisting of six short observations separated by significant gaps (see Table~\ref{tab:chandra_observations}, Figure~\ref{fig:lc}), there is generally a lack of rich X-ray photon statistics necessary for detailed time-resolved flare analysis and modeling \citep[e.g.,][]{Getman2011,Getman2021b}. Additionally, information regarding the morphology and duration of the identified large X-ray flares is limited. To derive rough estimates of intrinsic X-ray flare properties, such as flare peak coronal plasma temperature ($kT$) and scale of flaring coronal loop ($L$), we reasonably assume that the December 2023 X-ray flares follow the trends observed in large COUP X-ray flares from the 2003 Chandra observations, as analyzed and modeled by \citet{Getman08a,Getman08b}. Figures~\ref{fig:flare_basics}(e,f) display these trends. The trends are modeled using B-spline quantile regression, which offers enhanced flexibility for capturing non-linear relationships \citep{He1999, Ng2007, Ng2020}.  

These trends suggest that, for example, an X-ray super-flare with an energy of $\log(E_{X,fl}) \sim 35$~erg exhibits median values and interquartile ranges (IQRs) for the flare peak plasma temperature and loop scale as follows: median $kT \sim 3.5$~keV, $IQR_{kT} \sim (2.7, 4.5)$~keV, median $L \sim 2.4$~R$_{\odot}$, and $IQR_L \sim (1.6, 3.5)$~R$_{\odot}$. For an X-ray mega-flare with an energy of $\log(E_{X,fl}) \sim 36.5$~erg, these values are significantly higher: median $kT \sim 9$~keV, $IQR_{kT} \sim (6, 21)$~keV, median $L \sim 11$~R$_{\odot}$, and $IQR_L \sim (6, 20)$~R$_{\odot}$. Note that the $kT$ and $L$ values could still be underestimated because the inferred X-ray flare energy values for the December 2023 flares often represent lower limits rather than true energy values.

Nevertheless, these estimates of flare quantities could prove valuable for comparison with empirical measurements of stellar magnetic field strengths from HPF, H$^{13}$CO$^{+}$ disk fluxes from ALMA, CME radio fluxes from VLBA, and for future modeling efforts involving magnetic dynamos, disks, and CMEs.

Out of 245 young X-ray-emitting stars, 81 exhibiting notable X-ray flares in December 2023 have been targeted by followup obervations either with VLBA, ALMA, or HET-HPF. Table~\ref{tab:vlba_alma_het_targets} presents the properties of these 81 stars along with the inferred energetics of their X-ray flares. Among them, four stars (COUP~881, 1333, 1424, and 1463) were observed with HET-HPF; four (COUP~414, 561, 1174, and 1333) were observed with ALMA. 80 out of 81 stars were observed using VLBA. Figure~\ref{fig:lc_12_examples} presents a dozen examples of X-ray light curves for all the HET-HPF and ALMA targets, along with additional VLBA targets featuring some of the most powerful X-ray flares. Recall that detailed X-ray photon arrival diagrams and light curves for all the stars examined in this study are available in the electronic Figure Sets \ref{fig:phot_arrival} and \ref{fig:lc}.

{\it HET-HPF targets.} HPF is demonstrated effective for Zeeman measurements. For instance, 0.8-1.7~kG stellar surface magnetic field strengths are readily seen in exposures of the $r = 14.1$~mag M7-type Teegarden's Star. Similar detections were also made in Barnard's Star (M4-type) and AD Leo (M3-type). All of them from the broadening of the 1243~nm K I absorption line \citep{Terrien2022}. The detection of $B_{spot} = (10-20)$~kG fields expected in Orion Nebula super- and mega-flaring stars (\S~\ref{sec:intro}) should be achieved  using the magnetic intensification technique relying on Ti lines at $960-980$~nm with different magnetic sensitivity \citep{Kochukhov2021}. 

The strengths of Ti lines peak in late-K and M-type stars. Additionally, stars should be NIR bright ($J<11.5$~mag) to achieve a reasonable signal-to-noise ratio (SNR) in the spectrum. Ensuring accurate measurements of magnetic strengths necessitates that the lines are not significantly broadened due to stellar rotation or veiled by excess continuum emission due to disk accretion. To adhere to these constraints, we selected four X-ray flaring stars (COUP~881, 1333, 1424, and 1463) as targets for HPF observations.

All four are NIR bright (with $J$ magnitudes ranging from 10.8 to 11.5) stars of early M-type with masses ranging from 0.6 to 0.9~M$_{\odot}$. These stars demonstrate relatively slow rotation, with projected rotational velocities ranging from 9 to 14~km~s$^{-1}$. Notably, two of these stars (COUP~1333 and 1463) are surrounded by protoplanetary disks; however, they exhibit low accretion rates, as reported by \citet{Getman05}. All of them are subject to low dust extinction, with $A_V$ values not exceeding 2 magnitudes, indicating their probable membership in the ONC cluster \citep{Feigelson05}. Two stars (COUP~881, 1463) exhibited exceptionally strong X-ray super-flares ($\log(E_{X,fl}) > 35.3$~erg), while the other two (COUP1333, 1424) generated even more powerful mega-flares ($\log(E_{X,fl}) > 36.1$~erg; see Figure~\ref{fig:lc_12_examples}).

{\it ALMA targets.} Our ALMA observations are focused on detecting and monitoring optically thin isotopologues of HCO$^{+}$, such as H$^{13}$CO$^{+}$ and HC$^{18}$O$^{+}$, to investigate the theoretically predicted temporal enhancements of HCO$^{+}$ resulting from the disk's response to X-ray flares and changes in its chemistry \citep{Waggoner2023}.

The selection of ALMA targets was based on several criteria: previous detection of mm-band dust continuum emission to ensure the presence of massive protoplanetary disks, previous observations of gas emission indicating a high abundance of HCO$^{+}$, minimal contamination from surrounding parental molecular cloud material, and visual source extinction below 15 magnitudes to guarantee reliable spectral measurements.

The four ALMA targets selected, namely COUP~414, 561, 1174, and 1333, exhibit a range of spectral types from early M-type to G-K. They vary widely in terms of stellar properties, encompassing a broad spectrum of stellar mass (ranging from 0.5 to 3.6 M$_{\odot}$), source extinction (ranging from 0 to 14 magnitudes), and SED IRAC slope (ranging from -1.5 to 0.3). Of these targets, COUP~1333, exhibiting mega-flaring activity in X-rays, is the sole target observed by both HET-HPF and ALMA. In contrast, the remaining three ALMA targets showed more modest X-ray super-flares, with log X-ray energies ranging from 34.4 to 35.0 erg.

{\it VLBA targets.} Many of the 80 X-ray flaring stars observed by VLBA in December 2023 (Table~\ref{tab:vlba_alma_het_targets}) have been previously detected in radio surveys of the region \citep{Forbrich2021}. This prior detection enables precise alignment of source astrometry, facilitating the detection of potential shifts in the centroids of stars caused by X-ray flare-related CMEs moving away from the stars (\S~\ref{sec:intro}).

\begin{figure*}[h]
\centering
\includegraphics[width=0.95\textwidth]{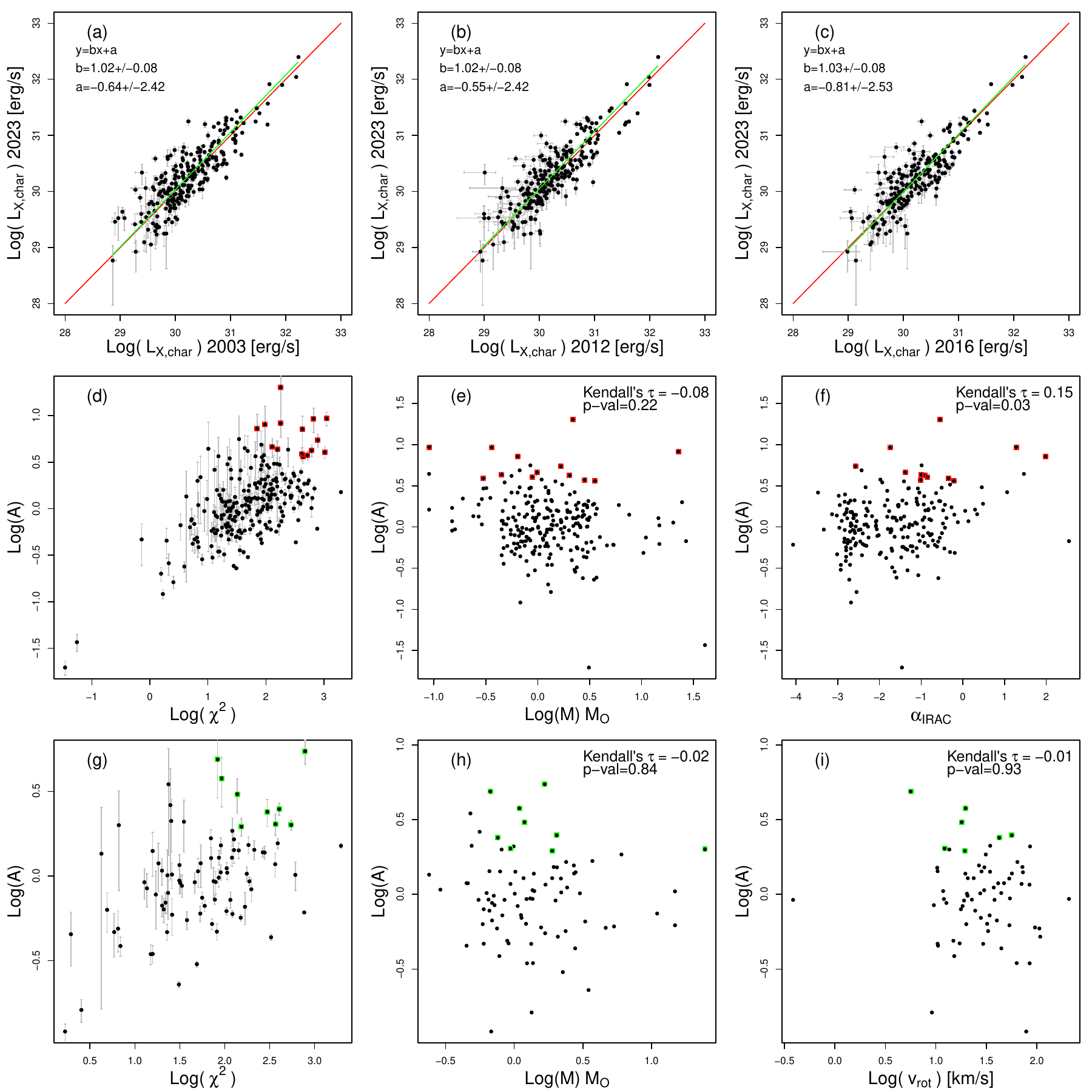}
\caption{(a-c) Comparison of stellar X-ray characteristic luminosities across different epochs. The unity lines are depicted in red. Linear regression fits using the Major Axis algorithm within R's {\it lmodel2} function are shown in green. Figure legends list the intercept, slope, and half-widths of the 95\% confidence intervals inferred by {\it lmodel2}. (d-f) Long-term X-ray characteristic variability amplitude ($A$) plotted against $\chi^2$, stellar mass, and SED IRAC slope for the entire stellar sample. Stars with the highest statistically reliable $A$ are highlighted with red points. Figure legends provide coefficients and p-values from Kendall's $\tau$ test. (g-h) Similar to (d-e), but for the subset of diskless stars. Diskless stars with the highest statistically reliable $A$ are denoted with green points. (i) For diskless stars, amplitude $A$ plotted against stellar rotation velocity.} \label{fig:cyclic_activity}
\end{figure*}

These 80 stars exhibit a wide range of PMS stellar characteristics (Table~\ref{tab:vlba_alma_het_targets}). Estimates are available for 96\% of their spectral types, 96\% of their masses, 85\% of their IRAC SED slopes, and 61\% of their rotation quantities (either periods or projected rotation velocities). Regarding spectral types, 45\% are classified as M-type, 46\% as K-type, 5\% as G-type, and 4\% as B-type stars, respectively. The stellar masses in the sample range between 0.1 and 24~M$_{\odot}$, with a median value of 0.9~M$_{\odot}$. Similarly, stellar radii span from 1 to 8~R$_{\odot}$, with a median value of 2.3~R$_{\odot}$. Source extinctions range from 0 to 36~mag, with a median value of 4.6~mag. SED IRAC slopes span from -3.5 to 1.5, with a median value of -1.5. Time-averaged COUP X-ray luminosities span from $8.7 \times 10^{28}$ to $5.7 \times 10^{31}$~erg~s$^{-1}$, with a median value of $1.8 \times 10^{30}$~erg~s$^{-1}$. The energies of the large X-ray flares range from $6.5 \times 10^{33}$ to $8.4 \times 10^{36}$~erg, with a median value of $2.4 \times 10^{35}$~erg.

Figure~\ref{fig:lc_12_examples} showcases five among the most powerful X-ray mega-flares with $\log(E_X) > 36$~erg produced by the VLBA target-stars COUP~539, 871, 939, 1333, and 1424. The apparent durations of these extraordinary flares easily exceed 100~ksec, and the apparent amplitudes (as ratios of the flare peak X-ray luminosity to the characteristic X-ray luminosity) exceed a factor of 9 for all and 40 for three of these flares.

\begin{deluxetable*}{ccccccccccccc}
\tabletypesize{\normalsize}
\tablecaption{Candidates for Future Cyclic Activity Analyses \label{tab:cyclic_candidates}}
\tablewidth{0pt}
\tablehead{
\colhead{Src.} & \colhead{$\chi^2$} & \colhead{$A$} & \colhead{$\sigma$\_$A$} &
\colhead{SpT} & \colhead{$A_V$} & \colhead{$M$} & \colhead{$R$} & \colhead{$\alpha_{IRAC}$} & \colhead{$P_{rot}$} & \colhead{$v_{rot} sin(i)$} & \colhead{$ME$} & \colhead{$\log(L_{X})$} \\
\colhead{} & \colhead{} & \colhead{} & \colhead{} &  \colhead{} & \colhead{(mag)} & \colhead{(M$_{\odot}$)} & \colhead{(R$_{\odot}$)} & \colhead{} & \colhead{(day)} & \colhead{(km~s$^{-1}$)} & \colhead{(keV)} & \colhead{(erg~s$^{-1}$)} \\
\colhead{(1)} & \colhead{(2)} & \colhead{(3)} & \colhead{(4)} & \colhead{(5)} & \colhead{(6)} & \colhead{(7)} & \colhead{(8)} & \colhead{(9)} & \colhead{(10)} & \colhead{(11)} & \colhead{(12)} & \colhead{(13)}
}
\startdata
COUP997 & 783.2 & 5.4  & 0.9 & K8 & 4.6 & 1.7 & 2.2 & -2.6 & \nodata & \nodata & 1.5 & 30.8 \\
COUP1087 & 403.4 & 2.5 & 0.2 & G/K & 3.0 & 2.0 & 3.3 & -2.8 & \nodata & 56.5 & 1.3 & 31.4 \\
COUP1127 & 137.7 & 3.0 & 0.7 & K7: & 4.5 & 1.2 & 1.9 & -2.3 & 7.2 & 18.0 & 1.5 & 30.4 \\
COUP1151 & 153.2 & 1.9 & 0.2 & K6 & 3.5 & 1.9 & 3.2 & -2.7 & 11.5 & 19.4 & 1.3 & 30.8 \\
COUP1210 & 92.5 & 3.8  & 1.2 & K2/4 & 5.5 & 1.1 & 2.5 & -2.0 & \nodata & 19.7 & 1.6 & 30.5 \\
COUP1232 & 547.7 & 2.0 & 0.1 & B1:V & 2.5 & 24.5 & 7.9 & -2.4 & \nodata & \nodata & 0.9 & 31.8 \\
COUP1355 & 83.2 & 4.9  & 2.0 & M3.5 & 0.0 & 0.7 & 1.3 & -2.4 & 10.4 & 5.6 & 1.2 & 30.1 \\
COUP1424 & 366.1 & 2.0 & 0.3 & M1 & 2.1 & 0.9 & 1.9 & -2.8 & 10.6 & 12.2 & 1.3 & 30.3 \\
COUP1516 & 297.5 & 2.4 & 0.4 & K1/4 & 0.0 & 0.8 & 1.4 & -3.0 & \nodata & 42.4 & 1.1 & 30.2 \\
\enddata 
\tablecomments{A subset of 9 diskless PMS stars exhibiting the highest statistically reliable amplitudes of long-term X-ray characteristic variations (green points in Figures~\ref{fig:cyclic_activity}(g-i)). Column 1: Source name. Columns 2-4: $\chi^2$, amplitude of variation and its 1-$\sigma$ uncertainty. Columns 5-13: Stellar properties including spectral type, source's visual extinction, stellar mass and radius, SED IRAC slope, stellar rotation period and projected velocity, X-ray median energy, and time-averaged X-ray luminosity (all from Table~\ref{tab:stellar_props}).
}
\end{deluxetable*}

\subsection{Search for signs of magnetic cyclic activity} \label{sec:seach_for_cyclic_activity} Figures~\ref{fig:cyclic_activity}(a-c) display the comparison of stellar X-ray characteristic luminosities across different epochs. The unity line falls within the confidence intervals of the model fits for all multi-epoch comparisons. These multi-epoch bivariate distributions exhibit no biases and IQR dispersions of 0.4~dex. The lack of strong biases indicates that our X-ray data analyses are accurate.

For each star, we calculate a simple  $\chi^2$ statistic to measure the deviation of its characteristic X-ray luminosities for each available epoch ($L_{X,char,i}$) from the characteristic luminosity averaged across all epochs ($\langle L_{X,char} \rangle$), considering the uncertainties associated with each epoch $i$, $\epsilon L_{X,char,i}$:
\begin{equation}
\chi^2 = \sum_{i = 1}^4 {\left(\frac{(L_{X,char,i} - \langle L_{X,char} \rangle)^2}{\epsilon  L_{X,char,i}^2}\right)}.
\end{equation}
We also compute the amplitude of the long-term variations in the X-ray characteristic luminosity as:
\begin{equation}
A = \frac{{\text{max}(L_{X,\text{char}}) - \text{min}(L_{X,\text{char}})}}{{\text{min}(L_{X,\text{char}})}}
\end{equation}

From the plot of $A$ as a function of $\chi^2$ (Figure~\ref{fig:cyclic_activity}d), we select 15 stars with the highest statistically reliable amplitudes (red points). These stars are COUP~315, 403, 444, 504, 536, 539, 599, 621, 801, 823, 901, 997, 1054, 1139, and 1298. Figure~\ref{fig:cyclic_activity}(e) indicates that the amplitudes $A$ of the full PMS stellar sample do not correlate with stellar mass or rotation (the latter graph is not shown). However, we observe a marginally statistically significant correlation with the indicator of disks (Figure~\ref{fig:cyclic_activity}f), and the sample of 15 stars with the highest amplitudes (Figure~\ref{fig:cyclic_activity}d) primarily comprises disk-bearing stars (except for COUP~997). Their source extinctions are generally systematically higher than those of the entire 245-star sample (graph is not shown). The presence of disks suggests that their long-term variability may not stem from magnetic cyclic activity but rather, possibly, from long-term variations in accretion columns \citep{Cody2022}. These accretion columns are expected to attenuate stellar X-ray emission, either due to the presence of dense, cold, non-X-ray-emitting gas or due to additional gas absorption within these columns \citep{Preibisch05}.

To mitigate possible long-term effects of disks and accretion, our subsequent step involved analyzing a subset of stars consisting only of those without disks (Figures~\ref{fig:cyclic_activity}(g-i)). Within this subset, we observed that the X-ray characteristic variability amplitudes exhibited no correlations with either stellar mass or rotation. Among them, nine stars (highlighted in green) demonstrated statistically reliable, highest X-ray characteristic variability amplitudes (see Table~\ref{tab:cyclic_candidates}). Among these, one system, COUP~1232, is a high-mass hierarchical triple \citep{Stelzer05}. In such case, the characteristic variability may be unrelated to magnetic cyclic activity. The remaining eight stars require further analysis utilizing extended archival {\it Chandra} ACIS-HETG data, a task that lies beyond the scope of the current paper.

In our analysis, considering the uncertainties associated with their amplitudes $A$, we observe that the majority of PMS stars display X-ray characteristic variability amplitudes below 1.5. This variation is notably smaller than the factor of $>10$ difference seen between the maximum and minimum X-ray activity levels on the current Sun, attributed to its 11-year cyclic activity \citep{Judge2004}. 

For the sample of 87 diskless PMS stars shown in Figure \ref{fig:cyclic_activity}g, we performed simulations to derive the cumulative distribution functions (CDFs) of their apparent X-ray variability amplitude $A_{app}$. These simulations were conducted with various assumptions about the intrinsic cyclical activity periods ($P_{cyc}$) and intrinsic cyclical amplitudes ($A_{intr}$). Each star's X-ray cyclical activity was modeled with a sinusoidal pattern and a random phase. The simulated stars were observed four times, with observation gaps reflecting the actual gaps between the four observation epochs listed in Table~\ref{tab:chandra_observations}. Simulations covered a broad range of $A_{intr}$ and three different $P_{cyc}$ values (1, 5, and 10 years). The real observed ECDF of $A$ (shown in Figure~\ref{fig:cyclic_activity}g) was compared to the simulated CDFs of $A_{app}$ using the Anderson-Darling test to infer the properties of the intrinsic cyclical activity of our 87 diskless stars.

\begin{figure}[h]
    \centering
     \includegraphics[width=0.48\textwidth]{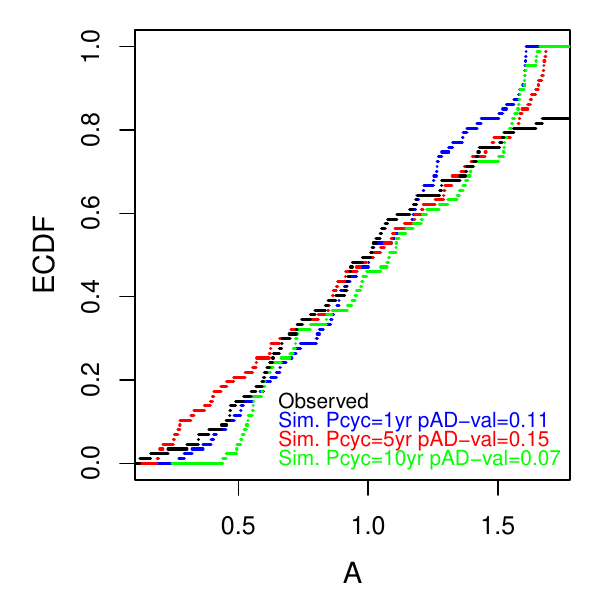}
    \caption{Comparison of the observed X-ray characteristic variability amplitude $A$ (black) for our sample of 87 diskless PMS stars (shown in Figure~\ref{fig:cyclic_activity}g) with the $A_{app}$ results from simulations (described in the text) assuming intrinsic cyclic activity periods of 1 year (blue), 5 years (red), and 10 years (green) and an intrinsic variability amplitude $A_{intr} = 1.7$. The p-values from the Anderson-Darling test for these comparisons are provided in the figure legend.} \label{fig:cyclic_activity_sumulations}
\end{figure}

The simulations indicate that an intrinsic cyclical activity amplitude of $A_{intr} = 1.7$, regardless of $P_{cyc}$, best matches the observed pattern of $A$. Figure~\ref{fig:cyclic_activity_sumulations} shows typical instances of $A_{app}$ CDFs from simulations with $A_{intr} = 1.7$ and $P_{cyc}$ values of 1, 5, and 10 years. Solutions with intrinsic amplitudes lower (higher) than 1.7 (not shown in the graph) would systematically fall to the left (right) of the observed $A$ locus.

Observed cases with $A > 1.6$, which represent less than 20\% of our stellar sample, deviate from the simulation outcomes (Figure~\ref{fig:cyclic_activity_sumulations}). For some of these cases, with $\log(\chi^2) < 2$ and large error bars on $A$ (Figure~\ref{fig:cyclic_activity}g), the $A$ estimates may be highly uncertain. For the remaining nine stars with the highest reliable $A$ values, listed in Table~\ref{tab:cyclic_candidates} and considered good candidates for future cyclic activity analyses, their intrinsic $A_{intr}$ may be systematically higher than the lower value of $A_{intr} = 1.7$ that fits the majority of our sample.

The diminished levels of long-term X-ray characteristic variability in most of our PMS stars align with the concept that young fully-convective stars possess saturated $\alpha^2$ magnetic dynamos, which operate independently of rotation \citep{Reiners2022}. Additionally, the presence of extensive surface coverage of coronal active regions and extremely elongated X-ray coronal structures \citep{Getman22,Coffaro2022,Getman2023} contributes to this observation. While PMS magnetic cycling is theoretically predicted \citep{Emeriau-Viard2017}, the existence of saturated X-ray coronal structures might mitigate the observable manifestation of PMS dynamo cycling. Furthermore, the presence of a saturated $\alpha^2$ dynamo could potentially inhibit strong magnetic cycling altogether.

\subsection{Do all PMS stars produce mega-flares?} \label{sec:flaring_restricted_vs_ubiq}

An important question regarding disk and planetary irradiation is whether powerful flares occur in only a subset of PMS stars or if the entire population produces mega-flares. Previous studies on powerful PMS flares \citep[e.g.,][]{Wolk05,Favata2005,Stelzer07,Caramazza07,Colombo07,Getman08a,Flaccomio2018,Getman2021} have been unable to address this question because the recurrence timescale of mega-flares is often longer than typical X-ray observation exposure times. However, we can examine this issue here, where over 200 stars exhibited flares in certain $Chandra$ exposures spanning two decades (see Table~\ref{tab:chandra_observations}). 

\begin{deluxetable}{cc | cc | c}[h]
\tabletypesize{\normalsize}
\tablecaption{Mega-flaring Stars in COUP vs. Later Epochs}  \label{tab:flare_contingency}
\tablewidth{0pt}
\tablehead{}
\startdata
 && \multicolumn{2}{c|}{Early Mega-flares} & \\ 
 &&  Y & N & Totals \\ \hline
Later  & Y &  9 &   9 &  ~18 \\ 
Mega-flares & N & 71 & 133 & 204 \\ \hline
       & Totals & 80 &  142 &  222 \\ 
\enddata
\end{deluxetable}

Ignoring the massive COUP~809 and COUP~1232 stars (O-type Theta~Ori~1C and Theta$^2$~Ori~A), we consider here a sub-sample of 222 stars, for which X-ray flares are registered in at least one of the COUP (i.e., epoch 2003 observations) observations (Table~\ref{tab:flare_energetics}). In this sample, 80 stars exhibited mega-flares during the lengthy COUP observation in 2003 (see Column ``Early Mega-flares Y'' in Table~\ref{tab:flare_contingency}), 5 mega-flared in 2012, 5 mega-flared in 2016, and 8 mega-flared in 2023 (see Row ``Later Mega-flares Y'' in Table~\ref{tab:flare_contingency}).  The latter 18 stars were all different; that is, no star exhibited a mega-flare twice in the 2012-2016-2023 epochs.  By itself, this does not suggest that mega-flaring is restricted to a small fraction of ONC stars.  Nine of these 18 stars also showed mega-flaring in the early COUP data (see the first cell with Column ``Early Mega-flares Y'' and Row ``Later Mega-flares Y'' in Table~\ref{tab:flare_contingency}). 

The question can be quantitatively examined using a $2 \times 2$ contingency table that compares the (non)flaring stars from the Early COUP observation with the (non)flaring stars from the Later epochs (Table~\ref{tab:flare_contingency}). Using the Fisher Exact Test\footnote{The Fisher Exact Test give probabilities of class differences in $2 \times 2$ contingency tables when counts are small \citep{Agresti06}.  Software implementations include $fisher.test$ in R and $scipy.stats.fisher\_exact$ in Python.} for count data, the probability that a correlation exists between the occurrence of Early and Later mega-flares is $p \simeq  0.2$.  This test would be significant with $p<0.01$ if only $\geq$12 of the 18 Late flaring stars were also Early flaring stars. There is thus no statistical evidence that the mega-flares repeatedly occur in the same sub-sample of Orion Nebula stars.

\begin{figure}[h]
    \centering
     \includegraphics[width=0.48\textwidth]{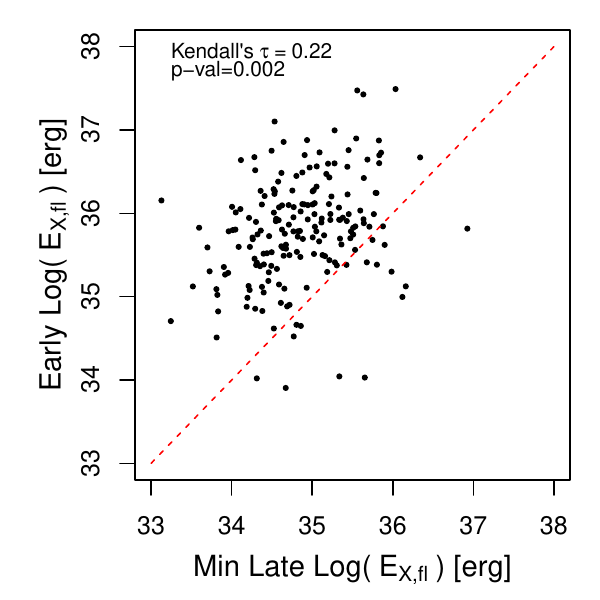}
    \caption{Bivariate relationship between Early (epoch 2003) and Late (epochs 2012, 2016, and 2023) flares for Orion Nebula PMS stars. If multiple flares are seen in more than one Later epoch, the weaker flare is shown. The dashed red line shows equal luminosities in Early and Late epochs.} \label{fig:early_later_flares}
\end{figure}

But we must recognize that the majority (133/222 = 60\% in Table~\ref{tab:flare_contingency}) of stars examined here did not exhibit mega-flares in any epoch, and most stars are too faint to appear at all in the shorter Later exposures.  There are two plausible explanations for this: (a) the exposures were not long enough and these stars would produce mega-flares if we looked long enough with $Chandra$ \citep{Getman2021}; or (b) these stars would never mega-flare so the incidence of mega-flaring is bimodal.   
However, a continuous unimodal population of flaring stars is suggested by the comparison of Early and Late flare energies in Figure~\ref{fig:flare_basics} and Figure~\ref{fig:early_later_flares}. Early and Late flare energies are associated with each other with a p-value of $0.002$, indicating statistical significance according to the nonparametric Kendall's $\tau$ correlation test, which is astrophysically attributed to the underlying correlation between flare energy and stellar mass/size (\S~\ref{sec:flares_props}). Again, as with Table~\ref{tab:flare_contingency}, there is no indication that the mega-flaring stars represent a special sub-population of PMS stars.  
 
Our findings suggest that all fully convective PMS stars produce extremely powerful X-ray flares. A solar-mass PMS star is expected to generate a few mega-flares per year \citep{Getman2021}. This would mean that every protoplanetary disk, and when they are dissipated \citep{Richert18}, every nascent primordial planetary atmosphere, is subject to millions of violent stellar magnetic reconnection events possibly including powerful CMEs (\S~\ref{sec:intro}).


\section{Conclusions} \label{sec:conclusions}
This study constitutes an important component of a comprehensive, multi-telescope effort aimed at exploring various aspects of PMS X-ray emission. In December 2023, we conducted nearly simultaneous observations utilizing the {\it Chandra}, HET, ALMA, and VLBA telescopes, targeting the PMS stellar members within the Orion Nebula star-forming region. Additionally, we analyzed archival {\it Chandra} data spanning previous epochs in 2003, 2012, and 2016. Leveraging this rich dataset, our project is poised to achieve several key scientific objectives (\S~\ref{sec:intro}), including the assessment of PMS surface magnetic field strengths following a large X-ray flare through synergistic HET-HPF/{\it Chandra} observations, exploration of X-ray flare impacts on disk chemistry via ALMA/{\it Chandra} investigations, hunting for flare-associated coronal mass ejections through VLBA/{\it Chandra} observations, and scrutiny of the multi-epoch behaviors of both PMS X-ray characteristic (baseline) and flare emission utilizing {\it Chandra} data.

Our careful analysis of the multi-epoch {\it Chandra} data, making use of CIAO and Acis-Extract tools, led to the extraction and X-ray photometric characterization of over 800 young stellar members of the Orion Nebula (\S~\ref{sec:xray_data_reduction}). Among these, we identified 245 stars as the brightest in the 2023 epoch, and compiled lists of their stellar properties (Table~\ref{tab:stellar_props}). Additionally, we generated atlases of stellar X-ray photon arrival diagrams and light-curves (Figure Sets \ref{fig:phot_arrival} and \ref{fig:lc}), identified significant flare and baseline segments, and computed the intrinsic X-ray luminosity levels and energetics associated with these segments (\S~\ref{sec:flare_identification}).

In Section \ref{sec:flares_props}, we observe a robust positive correlation between the X-ray energies of PMS flares identified across multiple epochs and the stellar mass and size. This correlation is attributed to the positive dependence of the underlying convection-driven dynamo on the stellar volume. Notably, flare energies do not exhibit a dependency on the presence or absence of protoplanetary disks. This suggests that the solar-type flare mechanism, involving both X-ray loop footpoints anchored in the stellar surface, is operational in PMS stars. While similar results have been reported previously for flares with $\log(E_{X,fl}) > 35$~erg, our current findings represent the first empirical observation of such trends in numerous less powerful PMS flares.

In Section \ref{sec:stars_flares_for_het_alma_vlba}, we analyze 81 X-ray flaring stars selected as targets for our subsequent observations with HET, ALMA, and VLBA. We meticulously examine their stellar properties alongside the characteristics of their prominent X-ray flares, offering comprehensive insights. These details are tabulated, serving as crucial supplementary material for our ongoing analyses and forthcoming HET-HPF, ALMA, and VLBA papers.

In Section~\ref{sec:seach_for_cyclic_activity}, we observe that the majority of the analyzed PMS stars exhibit relatively minor long-term variations in their baseline X-ray emission. This suggests that either convection-driven dynamos, operating within these rapidly rotating stars, do not generate magnetic cycles, or the PMS X-ray emission originating from coronal structures --- potentially saturated across extensive stellar coronal volumes --- diminishes the manifestation of dynamo cycling. We compile a list of several diskless stars that display the highest multi-epoch baseline variations. Constituting only a small fraction of our entire stellar sample, these stars are considered prime candidates for future investigations into PMS magnetic dynamo cycles using additional extended archival {\it Chandra} ACIS-HETG data.

In Section~\ref{sec:flaring_restricted_vs_ubiq}, we discover that X-ray mega-flaring is ubiquitous and not limited to a specific subset of stars. This indicates that every protoplanetary disk --- and upon their dissipation, every emerging primordial planetary atmosphere --- is subjected to millions of intense stellar magnetic reconnection events, potentially including powerful coronal mass ejections.

In Appendix \S~\ref{sec:chandra_me_changes}, we  evaluate the temporal increase in the apparent {\it Chandra} X-ray median energies of young Orion Nebula stars due to {\it Chandra}'s sensitivity degradation.

\section{Acknowledgments}
We are grateful to the anonymous referee for providing thoughtful and helpful comments that improved the manuscript. We thank the Chandra Mission Operations group for scheduling the X-ray observations for December 2023. We thank Patrick Broos (Penn State) for his valuable consultations regarding the use of the Acis Extract software package. This project is supported by the SAO {\it Chandra} grant  GO3-24010X (K. Getman, Principal Investigator) and the {\it Chandra} ACIS Team contract SV4-74018 (G. Garmire \& E. Feigelson, Principal Investigators), issued by the {\it Chandra} X-ray Center, which is operated by the Smithsonian Astrophysical Observatory for and on behalf of NASA under contract NAS8-03060. The {\it Chandra} Guaranteed Time Observations (GTO) data used here and listed in \citet{Getman05} were selected by the ACIS Instrument Principal Investigator, Gordon P. Garmire, of the Huntingdon Institute for X-ray Astronomy, LLC, which is under contract to the Smithsonian Astrophysical Observatory; contract SV2-82024. O. Kochukhov acknowledges support by the Swedish Research Council (grant agreements no. 2019-03548 and 2023-03667). Support for C.J.L. was provided by NASA through the NASA Hubble Fellowship grant No. HST-HF2-51535.001-A awarded by the Space Telescope Science Institute, which is operated by the Association of Universities for Research in Astronomy, Inc., for NASA, under contract NAS5-26555. S.A.D. acknowledges the M2FINDERS project from the European Research
Council (ERC) under the European Union's Horizon 2020 research and innovation programme
(grant No 101018682). This paper employs a list of Chandra datasets, obtained by the Chandra X-ray Observatory, contained in~\dataset[DOI: 10.25574/cdc.285]{https://doi.org/10.25574/cdc.285}.

\vspace{5mm}
\facilities{CXO}

\software{R \citep{RCoreTeam20}, HEASOFT \citep{Heasoft2014}, CIAO \citep{Fruscione2006}, AE \citep{Broos2010,Broos2012}}

\appendix
\section{Temporal Increase in Apparent {\it Chandra} X-ray Median Energy Due to Sensitivity Degradation} \label{sec:chandra_me_changes}

\begin{figure*}[h]
\centering
\includegraphics[width=0.97\textwidth]{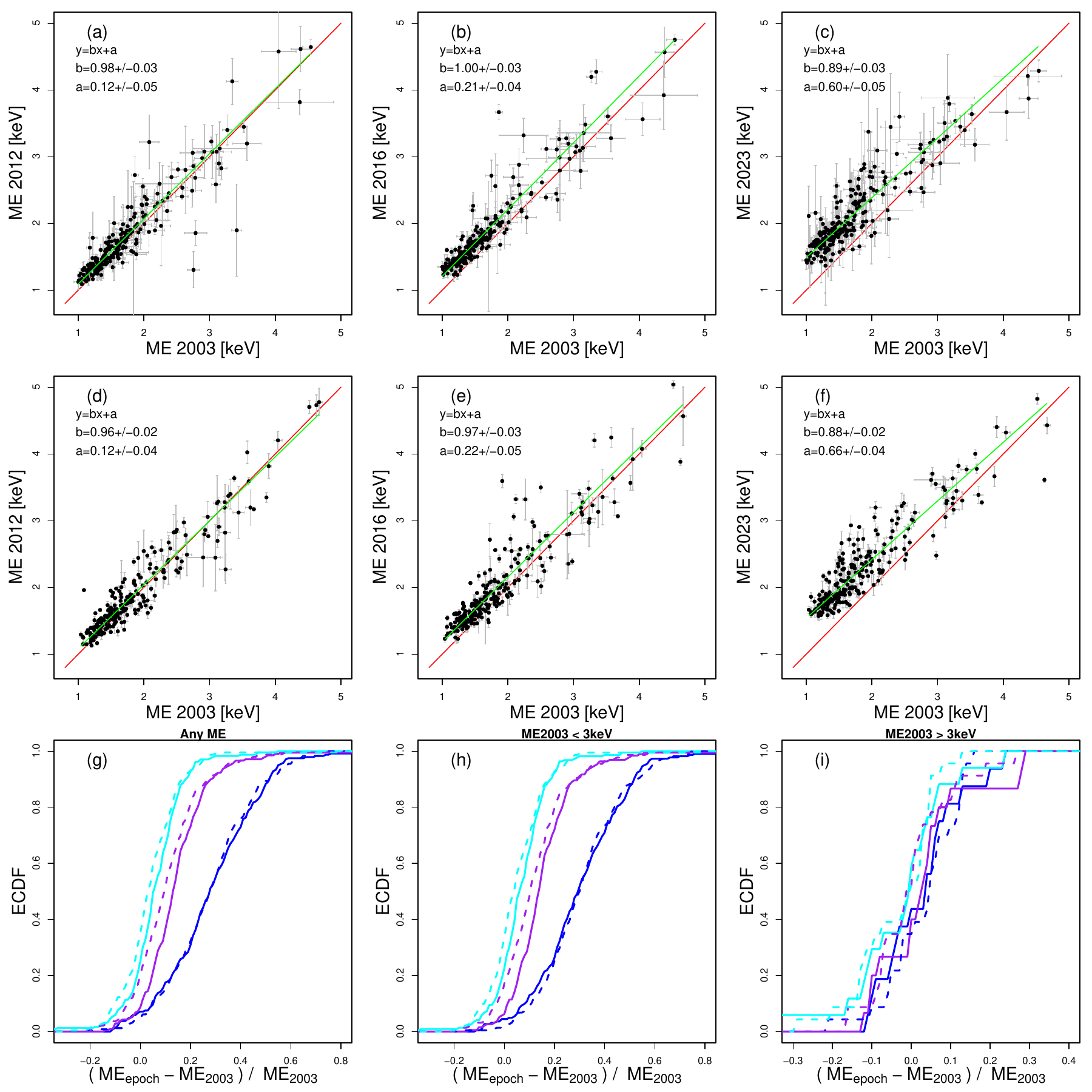}
\caption{(a-f) Comparison of \textit{Chandra}-ACIS-I apparent X-ray median energies across different epochs.  The red lines represent the unity reference, while the green lines show linear regression fits derived using the Major Axis algorithm, which treats variables symmetrically \citep{Warton2006}, from R's \textit{lmodel2} function \citep{LegeNume1998}. The figure legends display the intercept, slope, and 68\% confidence interval half-widths from the \textit{lmodel2} results. Panels (a-c) include only characteristic lightcurve segments (i.e., ``c'' in Column~9 of Table~\ref{tab:bb_segments}), while panels (d-f) encompass all lightcurve segments (i.e., large flares ``f'', mild variability ``n'', and characteristic ``c''). (g-i) ECDFs of fractional differences $(ME_{epoch} - ME_{2003}) / ME_{2003}$ across different energy ranges, epochs, and lightcurve segments. Energy ranges: (g) all stars, (h) stars with $ME_{2003} < 3$ keV, and (i) stars with $ME_{2003} \geq 3$ keV. Epochs: 2012 (cyan), 2016 (purple), and 2023 (blue). Lightcurve segments: solid curves represent only characteristic segments, while dashed curves represent all segments.} \label{fig:chandra_me_curves}
\end{figure*}

\newpage

\begin{deluxetable*}{ccccccccc}
\tabletypesize{\normalsize}
\tablecaption{{\it Chandra}-ACIS-I $ME$ Changes \label{tab:chandra_me_changes}}
\tablewidth{0pt}
\tablehead{
\colhead{Frac. Difference} & \colhead{Lc. Segments} & \colhead{Energy Range} &
\colhead{$25$\%} & \colhead{$50$\%} & \colhead{$75$\%} & 
\colhead{Intercept} & \colhead{Slope} & \colhead{p-val}\\
\colhead{(1)} & \colhead{(2)} & \colhead{(3)} & \colhead{(4)} & \colhead{(5)} & \colhead{(6)} & \colhead{(7)} & \colhead{(8)} & \colhead{(9)}
}
\startdata
$(ME_{2012} - ME_{2003}) / ME_{2003}$ & Char & Any $ME$  & 0.00 & 0.05 & 0.12 & $0.12 \pm 0.05$ & $0.98 \pm 0.03$ & 0.001\\
$(ME_{2016} - ME_{2003}) / ME_{2003}$ & Char & Any $ME$  & 0.06 & 0.13 & 0.20 & $0.21 \pm 0.04$ & $1.00 \pm 0.03$ & 0.001\\
$(ME_{2023} - ME_{2003}) / ME_{2003}$ & Char & Any $ME$  & 0.16 & 0.28 & 0.40 & $0.60 \pm 0.05$ & $0.89 \pm 0.03$ & 0.001\\
$(ME_{2012} - ME_{2003}) / ME_{2003}$ & All & Any $ME$  & -0.02 & 0.03 & 0.10 & $0.12 \pm 0.04$ & $0.96 \pm 0.02$ & 0.001\\
$(ME_{2016} - ME_{2003}) / ME_{2003}$ & All & Any $ME$  & 0.01 & 0.10 & 0.18 & $0.22 \pm 0.05$ & $0.97 \pm 0.03$ & 0.001\\
$(ME_{2023} - ME_{2003}) / ME_{2003}$ & All & Any $ME$  & 0.16 & 0.28 & 0.39 & $0.66 \pm 0.04$ & $0.88 \pm 0.02$ & 0.001\\
$(ME_{2012} - ME_{2003}) / ME_{2003}$ & Char & $ME_{2003} < 3$~keV & 0.01 & 0.06 & 0.12 & $0.05 \pm 0.06$ & $1.02 \pm 0.04$ & 0.001\\
$(ME_{2016} - ME_{2003}) / ME_{2003}$ & Char & $ME_{2003} < 3$~keV  & 0.07 & 0.13 & 0.21 & $0.08 \pm 0.06$ & $1.09 \pm 0.04$ & 0.001\\
$(ME_{2023} - ME_{2003}) / ME_{2003}$ & Char & $ME_{2003} < 3$~keV  & 0.20 & 0.29 & 0.42 & $0.35 \pm 0.08$ & $1.05 \pm 0.05$ & 0.001\\
$(ME_{2012} - ME_{2003}) / ME_{2003}$ & All & $ME_{2003} < 3$~keV  & -0.02 & 0.04 & 0.12 & $0.07 \pm 0.05$ & $1.00 \pm 0.03$ & 0.001\\
$(ME_{2016} - ME_{2003}) / ME_{2003}$ & All & $ME_{2003} < 3$~keV  & 0.03 & 0.10 & 0.18 & $0.02 \pm 0.08$ & $1.10 \pm 0.05$ & 0.001\\
$(ME_{2023} - ME_{2003}) / ME_{2003}$ & All & $ME_{2003} < 3$~keV  & 0.20 & 0.30 & 0.41 & $0.40 \pm 0.08$ & $1.05 \pm 0.05$ & 0.001\\
$(ME_{2012} - ME_{2003}) / ME_{2003}$ & Char & $ME_{2003} \geq 3$~keV  & -0.10 & -0.01 & 0.04 & $-2.45 \pm 1.85$ & $1.67 \pm 0.53$ & 0.003\\
$(ME_{2016} - ME_{2003}) / ME_{2003}$ & Char & $ME_{2003} \geq 3$~keV  & -0.04 & 0.02 & 0.05 & $-0.38 \pm 1.31$ & $1.14 \pm 0.37$ & 0.004\\
$(ME_{2023} - ME_{2003}) / ME_{2003}$ & Char & $ME_{2003} \geq 3$~keV  & -0.05 & 0.03 & 0.07 & $1.32 \pm 0.63$ & $0.64 \pm 0.18$ & 0.004\\
$(ME_{2012} - ME_{2003}) / ME_{2003}$ & All & $ME_{2003} \geq 3$~keV  & -0.12 & 0.00 & 0.03 & $-1.71 \pm 0.70$ & $1.44 \pm 0.20$ & 0.001\\
$(ME_{2016} - ME_{2003}) / ME_{2003}$ & All & $ME_{2003} \geq 3$~keV & -0.07 & -0.01 & 0.03 & $-0.67 \pm 0.92$ & $1.18 \pm 0.26$ & 0.001\\
$(ME_{2023} - ME_{2003}) / ME_{2003}$ & All & $ME_{2003} \geq 3$~keV  & -0.02 & 0.05 & 0.10 & $0.35 \pm 0.76$ & $0.92 \pm 0.21$ & 0.001\\
\enddata 
\tablecomments{Table of the 25th, 50th, and 75th percentiles of the $ME$ fractional difference distributions (as illustrated in Figure~\ref{fig:chandra_me_curves}(g-i)), along with the {\it lmodel2} fit results for the $ME_{epoch}$ versus $ME_{2003}$ relationships (as shown in Figure~\ref{fig:chandra_me_curves}(a-f)). Column~1: Form of fractional difference. Column~2: X-ray source lightcurve segments included in the calculations.  Column~3: Energy range.  Columns~4-6: Inferred 25th, 50th, and 75th percentiles of the distributions. Columns 7-9: Results from linear regression fits using {\it lmodel2}, including the intercept, slope, and p-value testing the null hypothesis of no relationship. All p-values are small, indicating statistically significant relationships between $ME_{epoch}$ and $ME_{2003}$.}
\end{deluxetable*}

\newpage

We assess the temporal evolution of the {\it Chandra}-ACIS-I apparent X-ray median energies for young stars in the Orion Nebula. These changes are primarily attributed to the accumulation of contamination on the optical blocking filters. The presented calculations may serve as a valuable resource for future researchers examining the {\it Chandra} archives.

Figure~\ref{fig:chandra_me_curves} compares median energy values across multiple epochs, accounting for various sets of X-ray lightcurve segments from Table~\ref{tab:bb_segments} and different energy ranges. Figures~\ref{fig:chandra_me_curves}
(a-f) confirm the expected linear trends between the X-ray median energy measured in 2003 and the median energies measured in the other three epochs, with systematically higher linear fit intercept values observed in the older epochs due to decreased \textit{Chandra} sensitivity. However, the trends are primarily influenced by the much more numerous X-ray softer stars. 

Figures~\ref{fig:chandra_me_curves}(g-i) and Table~\ref{tab:chandra_me_changes} offer a clearer representation of the actual distributions of median energy changes across different epochs, energy ranges, and selections of lightcurve segments.

Since large flares tend to produce a higher fraction of hard X-ray photons, and the longer 2003 {\it Chandra} exposure captures more large flares, the value of the fractional median energy $(ME_{\text{epoch}} - ME_{2003}) / ME_{2003}$ may be influenced by flares. The exclusive choice of characteristic lightcurve segments mitigates such potential effects. Indeed, Figures~\ref{fig:chandra_me_curves} (g,h) indicate small systematic differences between the fractional median energies for the same epoch, as shown by the shifts between the solid curves (using only characteristic lightcurve segments) and dashed curves (using all lightcurve segments) for the epochs 2012 and 2016. This suggests that for evaluating {\it Chandra} instrumental effects, it may be preferable to rely on results obtained from the selection of exclusively characteristic lightcurve segments.

The effects of instrumental sensitivity are significantly more pronounced in softer median energy ranges ($ME_{2003} < 3$~keV) compared to harder ranges ($ME_{2003} \geq 3$~keV). For example, when examining the $ME$ fractional difference between the epochs 2023 and 2003, the 50th and 75th percentiles of the difference distributions in softer energy ranges are 29\% and 42\%, respectively. In contrast, for harder energy ranges, these quantities are only 3\% and 7\%, respectively (Table~\ref{tab:chandra_me_changes}).

\bibliography{my_bibliography}{}
\bibliographystyle{aasjournal}

\end{document}